\documentclass[amsmath,amssymb,12pt,superscriptaddress,nofootinbib]{revtex4}

\usepackage[dvipdfmx]{graphicx}
\usepackage{dcolumn}
\usepackage{bm}
\usepackage{color}
\usepackage{enumitem}
\usepackage{tikz}
\usepackage{mathrsfs}
\usepackage[normalem]{ulem}
\setlistdepth{10}

\allowdisplaybreaks

\newcommand{\beq}{\begin{equation}}
\newcommand{\eeq}{\end{equation}}

\newtheorem{thm}{Theorem}

\definecolor{DarkBlue}{rgb}{0,0,0.7} 

\definecolor{DarkRed}{rgb}{0.65,0,0}

\begin{document}
\baselineskip5.5mm
\thispagestyle{empty}

{\baselineskip0pt
\small
\leftline{\baselineskip16pt\sl\vbox to0pt{
               \hbox{\it Division of Particle and Astrophysical Science, Nagoya University}
               \hbox{\it National Institute of Technology, Maizuru College}
                             \vss}}
\rightline{\baselineskip16pt\rm\vbox to20pt{
            {
            }
\vss}}%
}

\author{Masataka~Tsuchiya}\email{tsuchiya.masataka@h.mbox.nagoya-u.ac.jp}
\affiliation{
Division of Particle and Astrophysical Science,
Graduate School of Science, Nagoya University, 
Nagoya 464-8602, Japan
}

\author{Tsuyoshi~Houri}
\email{t.houri@maizuru-ct.ac.jp}
\affiliation{
National Institute of Technology,
Maizuru College,
Kyoto 625-8511, Japan
}

\author{Chul-Moon~Yoo}\email{yoo@gravity.phys.nagoya-u.ac.jp}
\affiliation{
Division of Particle and Astrophysical Science,
Graduate School of Science, Nagoya University, 
Nagoya 464-8602, Japan
}

\vskip2cm

\title{
The First Order Symmetry Operator on Gravitational Perturbations in the 5-dimensional Myers-Perry Spacetime \\
with Equal Angular Momenta}

\begin{abstract}
\vspace{5mm}
It has been revealed that the first order symmetry operator for the linearized Einstein equation on a vacuum spacetime can be constructed from a Killing-Yano 3-form. This might be used to construct all or part of solutions to the field equation. 
In this paper, we perform a mode decomposition of a metric perturbation on the Schwarzschild spacetime and the Myers-Perry spacetime with equal angular momenta in 5 dimensions, and investigate the action of the symmetry operator on specific modes concretely. 
We show that on such spacetimes, there is no transition between the modes of a metric perturbation by the action of the symmetry operator, and it ends up being the linear combination of the infinitesimal transformations of isometry.

\end{abstract}

\maketitle

\vspace{3cm}

\pagebreak

%

\section{Introduction}
\label{sec:I}

Hidden symmetry of spacetime, which is responsible for the integrability of various equations on a curved spacetime, has been studied actively in black hole physics. After hidden symmetry was recognized to be important in integrating the geodesic equation by separation of variables on the Kerr spacetime~\cite{PhysRev.174.1559,Carter:1968ks,Walker:1970un}, various field equations on various background spacetimes have been shown to be separable thanks to hidden symmetry.

It is well known that on the Kerr spacetime in 4 dimensions, the field equations for spin-$s$ fields for $s = 0$, $1/2$, $1$, $3/2$, $2$, namely, the Klein-Gordon equation, the Dirac equation, the Maxwell equation, the Rarita-Schwinger equation, and the linearized Einstein equation, are separable~\cite{
Teukolsky:1972my,Teukolsky:1973ha,Aksteiner:2016mol,Araneda:2016iwr,
Unruh:1973bda,Chandrasekhar:1976ap,Page:1976jj,
SilvaOrtigoza:1995wn,
Kegeles:1979an}. 
In contrast, on the Myers-Perry spacetime, which is a higher-dimensional generalization of the Kerr spacetime, the only spin-$s$ field equations for $s = 0$, $1/2$, $1$ are known to be separable; the separability of the equations for $s = 3/2$ and $2$ have not been revealed (see the magnificent wrap-up~\cite{Frolov:2017kze} of progress until 2017, references therein, and Refs.~\cite{Lunin:2017drx,Krtous:2018bvk,Frolov:2018ezx,Frolov:2018eza,Houri:2019lnu} for $s = 1$, published very recently). 
Separation of variables is also successful for $s = 1/2$ coupled with electromagnetic field in the 4-dimensional flat spacetime~\cite{Breev:2015aca}, $s = 1$ in the Pleba\'{n}ski-Demia\'{n}ski spacetime and the Wahlquist
spacetime~\cite{Frolov:2018pys,Houri:2019nun}, and differential $p$-forms for $p = 1$, $2$, $3$, $4$ in the Myers-Perry
spacetime (in any dimention for $p = 1$, 2, 3 and 10 dimension for $p = 4$)~\cite{Lunin:2019pwz}.

When a field equation is separable thanks to hidden symmetry, one can
find associated differential operators known as symmetry operators, which commute with the operator defining the field equation, and the separation constants are obtained as the eigenvalues of the symmetry operators. 
In this manner, hidden symmetry, the separability of field equations, and the existence of symmetry operators are closely related to each other.

Symmetry operators are also useful tools for obtaining all or part of solutions to field equations by acting them on a solution to obtain another solution like ladder operators as creation/annihilation operators in quantum mechanics. 
Recently, the following ladder operators have been reported: the
spin ladder operators between $s = p$ and $s = p + 1/2$ with $p = 0$, 1~\cite{Acik:2017epj}, and
the mass ladder operator for $s = 0$~\cite{Cardoso:2017qmj,Cardoso:2017egd,Muck:2017vzb}.

In Refs.~\cite{Michishita:2018auz,Michishita:2019eeh,Michishita:2020umh}, possible forms of first order symmetry operators for $s = 3/2$, $s = 2$ and some differential $p$-forms on a vacuum spacetime have
been investigated. 
From the series of results, one realizes that (conformal) Killing-Yano forms are responsible for the existence of first order symmetry operators for those equations. Particularly, in the $s = 2$ case, which we focus on in this paper, a Killing-Yano 3-form is responsible for the existence of the first order symmetry operator for the linearized Einstein equation.

Having the results in Refs.~\cite{Michishita:2018auz,Michishita:2019eeh,Michishita:2020umh}, one may be curious about the action of the symmetry operator, and it is very natural to think of the possibility of the symmetry operator to be a ladder operator. 
Specifically, one would also be interested in the commutation relations with other symmetry operators and the eigentensors of the symmetry operator. 
If the symmetry operator does not commute with the isometries that define the mode decomposition, 
the symmetry operator would be a certain combination of the raising and lowering operators.
Note that, although it is clear that a Killing-Yano 3-form does not provide both of the raising and lowering operators of first order, 
it does not necessarily imply that
the symmetry operator does not play the role of the ladder operator.
In order to grab how the symmetry operator acts on a metric perturbation, we consider a specific background metric and clarify the structure of the map of the symmetry operator in the space of the solutions to the linearized Einstein equation.

Recall that the odd-dimensional Myers-Perry spacetime admits odd rank
Killing-Yano forms. 
However, the master equations for the linearized Einstein equation on such spacetimes have not been clarified. 
As the simpler cases, we consider the Schwarzschild spacetime and the Myers-Perry spacetime with equal angular momenta in 5 dimensions. Since these spacetimes admit a Killing-Yano 3-form, we describe the map between linear metric perturbations by the symmetry operator constructed from the Killing-Yano 3-form by means of a certain mode decomposition of a metric perturbation.

In the Schwarzschild case, we first decompose a metric perturbation in terms of the scalar, vector and tensor harmonics on the 3-sphere~\cite{Ishibashi:2011ws}. 
Then, we further introduce more detailed classes of the harmonics, presented in Refs.~\cite{doi:10.1063/1.523649,Lindblom:2017maa}. 
Finally, the map of the symmetry operator is clearly described
in terms of these classes of the harmonics without using any field equation.

In the case of the finite angular momenta, obviously, the same decomposition of a metric perturbation as the Schwarzschild case is not available because of the reduced symmetry. 
We so rely on another way presented in Refs.~\cite{Murata:2007gv,Murata:2008yx}, where the mode decomposition is performed by the group theoretical method originally proposed in Ref.~\cite{Hu:1974hh}. 
We describe the map of the several modes dealt with in Ref.~\cite{Murata:2008yx} and simplify them by using the corresponding field equation.

Before summarizing the results in this paper, we point out possible nontrivial issues in the map of the symmetry operator. 
First, although we know that the map converts a solution into another solution, the preservation of the boundary condition is not trivial at a glance. 
Secondly, the preservation of the mode seems not necessarily to be guaranteed. That is, we need to check whether the map of the symmetry operator acts as a ladder operator or no, which is one of our main interests. 
In spite of these nontrivial issues, as is stated below, the symmetry operator we analyze results in the combination of the infinitesimal transformations of isometry, acting on each mode of a
metric perturbation.

In the Schwarzschild case, the symmetry operator acts as the Hodge star on the exterior derivative on the 3-sphere, $\hat \star\hat d$. 
There are three classes of vector harmonics and six classes of tensor harmonics on the 3-sphere having the same mode numbers, as is shown in Refs.~\cite{doi:10.1063/1.523649,Lindblom:2017maa}. 
The symmetry operator maps scalar harmonics and some classes of vector and tensor harmonics to zero, and the other classes are divided into pairs such that a class is mapped to the other. 
By considering the linear combination of the two classes in the
pair, we have eigenvectors and eigentensors of the symmetry operator.

In the finite angular momenta case, we first find that the symmetry operator causes no transition between two different modes for any rank-2 symmetric tensor. 
In other words, the result shows that the symmetry operator commutes with the Lie derivatives associated with the rotational isometries. 
Secondly, by evaluating the master variables of the modes dealt with
in Ref.~\cite{Murata:2008yx}, we find that the map of the symmetry operator results in the combination of identity and phase shift, where the phase shift is originated from the imaginary part of the frequency.

This paper is organized as follows. In Sec.~\ref{sec:II0}, we introduce the notion of symmetry operator for linear metric perturbation, the theorem on possible form of symmetry operators on vacuum spacetimes, presented in Ref.~\cite{Michishita:2019eeh},
and the metric and the Killing-Yano 3-form on the 5-dimensional Myers-Perry black hole spacetime. 
In Sec.~\ref{sec:II}, we apply the unique method of mode decomposition for warped product spaces in the Schwarzschild spacetime, and we describe the map in terms of scalar, vector and tensor harmonics. 
In Sec.~\ref{sec:III}, we apply the group theoretical method of mode decomposition in the 5-dimensional Myers-Perry spacetime with finite equal angular momenta.
We show that the symmetry operators constructed from the Killing-Yano 3-form and generators of $U(1) \times SU(2)$ are commuting with each other, and the symmetry operator constructed from the Killing-Yano 3-form results in the combination of the operators associated with the isometries with respect to several modes by applying the field equation in the final step. 
Sec.~\ref{sec:V} is devoted to summary and discussion.

\section{Killing-Yano 3-form and linear metric perturbations in the 5-dim. Myers-Perry spacetime with equal angular momenta}
\label{sec:II0}

\subsection{First order symmetry operators for linear metric perturbations}

Let ${\cal M}_{\kappa\tau}{}^{\mu\nu}$ be the second order operator in the field equation ${\cal M}_{\kappa\phi}{}^{\mu\nu}h_{\mu\nu}=0$ for a linear metric perturbation $g_{\mu\nu}\mapsto g_{\mu\nu}+h_{\mu\nu}$ of a spacetime equipped with metric $g_{\mu\nu}$.
The operators $({\cal Q}_{\lambda\rho}{}^{\kappa\tau},{\cal S}_{\kappa\tau}{}^{\mu\nu})$ are called the symmetry operators for ${\cal M}_{\kappa\phi}{}^{\mu\nu}$ if they satisfy
\begin{equation}
{\cal Q}_{\lambda\rho}{}^{\kappa\tau}{\cal M}_{\kappa\tau}{}^{\mu\nu}
-{\cal M}_{\lambda\rho}{}^{\kappa\tau}{\cal S}_{\kappa\tau}{}^{\mu\nu}=0.
\label{com}
\end{equation}
The image ${\cal S}_{\lambda\rho}{}^{\mu\nu}h_{\mu\nu}$ of a linear metric perturbation $h_{\mu\nu}$ fulfills the field equation as well as $h_{\mu\nu}$,
\begin{equation}
{\cal M}_{\lambda\rho}{}^{\kappa\tau}{\cal S}_{\kappa\tau}{}^{\mu\nu}h_{\mu\nu}={\cal Q}_{\lambda\rho}{}^{\kappa\tau}{\cal M}_{\kappa\phi}{}^{\mu\nu}h_{\mu\nu}=0.
\end{equation}
All the first symmetry operators for the linear metric perturbations of vacuum spacetimes have been revealed by Y.~Michishita in 2019,
%
\begin{thm}[General form of first order symmetry operators~\cite{Michishita:2019eeh}]
\label{thm1}
First order symmetry operators $({\cal Q}_{\lambda\rho}{}^{\kappa\tau},{\cal S}_{\kappa\tau}{}^{\mu\nu})$ for linear metric perturbations $h_{\lambda\rho}$ for spacetimes of arbitrary $D\ge4$ dimensions satisfying Einstein's field equation $G_{\mu\nu}+\Lambda g_{\mu\nu}=0$ must be given in the form
\begin{subequations}
\begin{align}
{\cal Q}_{\lambda\rho}{}^{\mu\nu}h_{\mu\nu} & = 
ch_{\lambda\rho}
 + 2f_{(\lambda}{}^{\kappa\tau}\nabla_{|\kappa|}h_{\rho)\tau}
 + F_{\lambda\rho}{}^\mu\nabla^\nu h_{\mu\nu}
 + \Big(
 \pounds_{K}h_{\lambda\rho}
 +\frac{2\nabla_\kappa K^\kappa}{D} h_{\lambda\rho}\Big),
\label{solQ}
\\
{\cal S}_{\lambda\rho}{}^{\mu\nu}h_{\mu\nu} & = 
ch_{\lambda\rho}
+ 2f_{(\lambda}{}^{\kappa\tau}\nabla_{|\kappa|}h_{\rho)\tau}
+ \nabla_{(\lambda}[H^{\mu\nu}{}_{\rho)}h_{\mu\nu}]
+\pounds_{K}h_{\lambda\rho},
\label{solS}
\end{align}
\end{subequations}
where $c$ is a constant,
$f_{\mu\nu\lambda}$ is a Killing-Yano 3-form, $H^{\mu\nu}{}_\lambda=H^{\nu\mu}{}_\lambda$ is an arbitrary tensor,
$F_{\lambda\rho}{}^\mu=F_{\rho\lambda}{}^\mu$ is an arbitrary tensor, and
$K^\mu$ is a homothetic vector.
\footnote{A homothetic Killing vector is a conformal Killing vector of constant divergence.}
In particular, $K^\mu$ is given by a Killing vector for $\Lambda\neq 0$.
\end{thm}
Ignoring the contribution from multiplication of amplitude by constant and gauge transformations, 
we see that non-trivial first order symmetry operators on vacuum spacetimes are provided by homothetic vectors and Killing-Yano 3-forms.

We focus on the symmetry operators constructed with the Killing-Yano 3-form, the second terms in the right-hand side of Eqs.~\eqref{solQ} and \eqref{solS},
\begin{equation}
{\cal Q}_{\lambda\rho}{}^{\mu\nu}h_{\mu\nu}
={\cal S}_{\lambda\rho}{}^{\mu\nu}h_{\mu\nu}  
=2f_{(\lambda}{}^{\kappa\tau}\nabla_{|\kappa|}h_{\rho)\tau},
\end{equation}
where Eq.~\eqref{com} results in
\begin{equation}
[{\cal S},{\cal M}]_{\lambda\rho}={\cal S}_{\lambda\rho}{}^{\kappa\tau}{\cal M}_{\kappa\tau}{}^{\mu\nu}
-{\cal M}_{\lambda\rho}{}^{\kappa\tau}{\cal S}_{\kappa\tau}{}^{\mu\nu}=0.
\end{equation}

The odd-dimensional Kerr-NUT-(A)dS spacetimes are known to admit  Killing-Yano 3-forms.
In this paper, we consider the 5-dimensional Myers-Perry black hole spacetime with equal angular momenta 
as the simplest model, and we investigate the map of linear metric perturbations by the first symmetry operator associated with the Killing-Yano 3-form in detail.
\footnote{
The dimension $D=5$ is special in taking the Schwarzschild limit for the following reason.
The Killing-Yano 3-form in the $(2n+1)$-dimensional Kerr-NUT-(A)dS geometry considered here is the Hodge dual of the $2(n-1)$-form $\wedge^{(n-1)}\mathfrak{h}$, where $\mathfrak{h}$ is the closed conformal Killing-Yano $2$-form.
In the Schwarzschild limit, the wedge product of $\mathfrak{h}$ vanishes for $n\ge 3$, i.e. $D=7,9,\cdots$, and simultaneously
the Killing-Yano 3-form vanishes.
On the other hand, $\wedge^{(n-1)}\mathfrak{h}$ and the Killing-Yano 3-form both survive for $n=2$, i.e. $D=5$, and the Killing-Yano 3-form behaves as the volume form of $S^{3}$ as we will show in Sec.~\ref{sec:II}.
}

\subsection{5-dim. Myers-Perry black hole spacetime 
with equal angular momenta\\
and Killing-Yano 3-form
}
\label{sec:IIB}

In the Boyer-Lindquist coordinates $(t,r,\theta,\phi,\psi)$, the metric of the Myers-Perry spacetime with equal angular momenta is given by
\begin{align}
g&
 =-dt^{2}+\frac{\Sigma}{\Delta_{r}}dr^{2}+\Sigma\left\{ d\theta^{2}+ \sin^{2}\theta \cos^{2}\theta \big(d\phi-d\psi\big)^{2}
+\big(\sin^{2}\theta d\phi +\cos^{2}\theta d\psi\big)^{2}\right\}\notag\\
&\qquad+\frac{2M}{\Sigma}\left\{dt-a\big(\sin^{2}\theta d\phi +\cos^{2}\theta d\psi\big)\right\}^{2},
\label{met:equal}
\end{align}
where
\begin{equation}
\Delta_r = \Sigma^{2}/r^2 -2M \, , \quad \Sigma = r^2 +a^2.
\end{equation}
The constant $M$ is the mass of the black hole, and the constant $a$ represents the angular momenta of the black hole.
In the orthonormal frame given by
\begin{subequations}
\begin{align}
e^{0}&=\sqrt{\frac{\Delta_r}{\Sigma}}\big\{dt -a(\sin^{2}\theta d\phi +\cos^{2}\theta d\psi)\big\},\label{e0:equal}\\
e^{1}&=\sqrt{\frac{\Sigma}{\Delta_r}}~dr, \\
e^{2}&=\sqrt{\Sigma}\ {}d\theta,\\
e^{3}&=\sqrt{\Sigma}\,\sin\theta\cos\theta\big( d\phi -d\psi\big),\\
e^{5}&=\frac{1}{r}\big\{-adt +\Sigma(\sin^{2}\theta d\phi +\cos^{2}\theta d\psi)\big\},
\label{e5:equal}
\end{align}
\end{subequations}
where $e^{0}$ is timelike, and the others are spacelike,
the Killing-Yano 3-form on the spacetime is given by
\begin{equation}
f=-a\,e^{0}\wedge e^{1}\wedge e^{5}+r\,e^{2}\wedge e^{3}\wedge e^{5}.
\label{KY3form:equal}
\end{equation}

We investigate the map of linear metric perturbations of the spacetime by the symmetry operator associated with the Killing-Yano 3-form for the Schwarzschild case and the finite angular momenta case in Sec.~\ref{sec:II} and Sec.~\ref{sec:III}, respectively.
Hereafter, for convenience, we denote the image ${\cal S}_{\lambda\rho}{}^{\mu\nu}h_{\mu\nu}$ of a rank-2 symmetric tensor $h_{\mu\nu}$ by the symmetry operator associated with $f_{\lambda}{}^{\kappa\tau}$ as
\begin{equation}
Z_{\lambda\rho}=2f_{(\lambda}{}^{\kappa\tau}\nabla_{|\kappa|}h_{\rho)\tau}
\label{solN}
\end{equation}
during the investigation of both the Schwarzschild case and the finite angular momenta case.
We note that $h_{\mu\nu}$ does not necessarily satisfy the field equation ${\cal M}_{\kappa\phi}{}^{\mu\nu}h_{\mu\nu}=0$ until we declare that.

\section{Schwarzschild case}
\label{sec:II}

\subsection{Metric tensor and Killing-Yano 3-form}

In the non-rotating limit $a\to0$ of Eqs.~\eqref{met:equal}--\eqref{KY3form:equal}, we have the Schwarzschild metric
\begin{equation}
g=-F(r)dt^{2}+F(r)^{-1}dr^{2}+r^{2}\left(d\theta^{2}+\sin^{2}\theta d\phi^{2}+\cos^{2}\theta d\psi^{2}\right),
\label{metSch}
\end{equation}
where $F(r)=1-2M/r^{2}$,
and we also have that the Killing-Yano 3-form is proportional to the volume form of the unit 3-sphere $S^{3}$,
\begin{equation}
f
=r^{4}\sin\theta\cos\theta~d\theta\wedge d\phi\wedge d\psi.
\label{KYSch}
\end{equation}
We introduce the notations $y^{a}=(t,r)$ and $z^{i}=(\theta,\phi,\psi)$ suitable for warped product spaces, and rewrite Eqs.~\eqref{metSch} and \eqref{KYSch} as
\begin{align}
g&=g_{ab}(y)d y^{a}d y^{b}+r^{2}(y)\gamma_{ij}(z)d z^{i}d z^{j},\\
f_{ijk}&=r^{4}\hat\varepsilon_{ijk},
\label{KYSch2}
\end{align}
where $\gamma_{ij}$ is the metric of the unit sphere $S^{3}$, and $\hat\varepsilon_{ijk}$ is the volume form 
associated with $\gamma_{ij}$.
Note that razing and lowering induces $i,j,\cdots$ will be performed with respect to $\gamma_{ij}$ in the following, not $g_{ij}=r^{2}\gamma_{ij}$.

\subsection{Investigation of the symmetry operator}

Substituting the Killing-Yano 3-form~\eqref{KYSch2} on the Schwarzschild spacetime into Eq.~\eqref{solN}, and applying the notation of a warped product space by rewriting $\nabla$ using the covariant derivative $\hat D_{i}$ associated with $\gamma_{ij}$, we have
\begin{subequations}
\begin{align}
Z_{ab}&=0,\\
Z_{ai}&=\hat\varepsilon_{i}{}^{kl}\hat D_{k}h_{al},\label{ai}\\
Z_{ij}&=2\hat\varepsilon_{(i}{}^{kl}\hat D_{|k|}h_{j)l}.\label{ij}
\end{align}
\end{subequations}
It is worth noting that the symmetry operator exactly acts as the operation $\hat \star\hat d$ in Eq.~\eqref{ai}, where $\hat d$ denotes the exterior derivative, and $\hat\star$ denotes the Hodge star operator on $(S^{3},\gamma)$.

We apply the decomposition method 
summarized in Ref.~\cite{Ishibashi:2011ws}.
A rank-2 symmetric tensor $h_{\lambda\rho}$ can be uniquely decomposed to the 
scalar perturbations $h_{ab}$, $h_{a}$, $h_{L}$ and $h_{T}$, 
vector perturbations $h_{T}\,{}_{ai}$ and $h_{T}\,{}_{i}$ and tensor perturbation $h_{T}\,{}_{ij}$
as follows:
\begin{subequations}
\begin{align}
h_{ab}&=h_{ab},\\
h_{ai} &= \hat{D}_ih_a + h_{T}\,{}_{ai}, 
\label{hai:decomp}
\\
h_{ij} &= h_T\,{}_{ij} + 2{\hat D}_{(i} h_T\,{}_{j)} + h_L \gamma_{ij}+ {\hat L}_{ij} h_T,
\label{hij:decomp}
\end{align}
\end{subequations}
where 
$\hat L_{ij}=\hat D_{i}\hat D_{j}-\frac{1}{3}\gamma_{ij}\hat\triangle$, 
and the subscript ``$\,{}_{T}\,$'' represents the transverse-traceless property,
\begin{equation}
{\hat D}^ih_{T}\,_{ai} =0,\quad{\hat D}^ih_T\,{}_{i}  = 0,\quad {\hat D}^jh_T\,{}_{ij} = h_T\,{}^i{}_{i} = 0.
\label{property:h}
\end{equation}
$Z_{\lambda\rho}$ can be uniquely decomposed likewise
as follows:
\begin{subequations}
\begin{align}
Z_{ai} &= \hat{D}_iZ_a + Z_{T}\,_{ai}, 
\label{Zai:decomp}
\\
Z_{ij} &= Z_T\,{}_{ij} + 2{\hat D}_{(i} Z_T\,{}_{j)} + Z_L \gamma_{ij}+ {\hat L}_{ij} Z_T,
\label{Zij:decomp}
\end{align}
\label{Z:decomp}
\end{subequations}
where
\begin{equation}
{\hat D}^iZ_{T}\,_{ai} = 0, \quad 
{\hat D}^iZ_T\,{}_{i}  = 0,\quad
{\hat D}^jZ_T\,{}_{ij} = Z_T\,{}^i{}_{i} = 0.
\label{property:Z}
\end{equation}
Here, let us give another decomposition of $Z_{\lambda\rho}$ based on Eqs.~\eqref{ai} and \eqref{ij}.
$Z_{\lambda\rho}$ is given as the summation of the maps of each term of $h_{\lambda\rho}$, 
which we can formally write as
\begin{subequations}
\begin{align}
Z_{ai}&=Z_{ai}[h_{a}]+Z_{ai}[h_{T}\,_{ai}],
\label{Zai:decomp2}\\
Z_{ij}&=Z_{ij}[h_{T}\,{}_{ij}]+Z_{ij}[h_{T}\,{}_{i}]+Z_{ij}[h_{L}]+Z_{ij}[h_{T}].
\label{Zij:decomp2}
\end{align}
\label{Z:decomp2}
\end{subequations}
We compare the two expressions of $Z_{\lambda\rho}$, Eqs.~\eqref{Z:decomp} and \eqref{Z:decomp2} 
with each other 
in Appendix~\ref{app:Sch}.
In summary, first, all the scalar perturbations do not contribute,
\begin{subequations}
\begin{gather}
Z_{a}=0,\quad Z_{L}=0,\quad Z_{T}=0,\label{sch:rel1}\\
Z_{ai}[h_{a}]=0,\quad Z_{ij}[h_{L}]=0,\quad Z_{ij}[h_{T}]=0,
\label{sch:rel2}
\end{gather}
\end{subequations}
and secondly, we have the following natural relations among the contributions from vector and tensor perturbations,
\begin{equation}
Z_{T}\,_{ai}=\hat\varepsilon_{i}{}^{kl}\hat D_{k}h_{T}\,_{al},\quad
Z_{T}\,{}_{j}=\hat\varepsilon_{j}{}^{kl}\hat D_{k}h_{T}\,{}_{l},\quad
Z_{T}\,{}_{ij}=2\hat\varepsilon_{(i}{}^{kl}\hat D_{|k|}h_{T}\,{}_{j)l}.
\label{sch:rel3}
\end{equation}
Readers may refer to Appendix~\ref{app:Sch} for the details.

As is performed in Ref.~\cite{Ishibashi:2011ws}, the contributions in $h_{\lambda\rho}$ from scalar, vector and tensor perturbations 
can further be decomposed into scalar, vector and tensor harmonics.
A scalar harmonics $\mathbb{S}$, a vector harmonics $\mathbb{V}_{i}$ and a tensor harmonics $\mathbb{T}_{ij}$ on $S^{3}$ are an eigenscalar, eigenvector and eigentensor of the connection Laplacian $\hat\triangle:=\hat D^{i}\hat D_{i}$, where $\mathbb{V}_{i}$ and $\mathbb{T}_{ij}$ obey $\hat D^{i}\,\mathbb{V}_{i}=0$ and $\hat D^{i}\,\mathbb{T}_{ij}=\mathbb{T}^{i}{}_{i}=0$, respectively.
The scalar harmonics $\mathbb{S}$ and the vector harmonics $\mathbb{V}_{i}$ define the vector $\mathbb{S}_{i}$ and the tensors $\mathbb{S}_{ij}$ and $\mathbb{V}_{ij}$ by
\begin{equation}
\mathbb{S}_{i}:=-\frac{1}{\lambda_{s}}\hat D_{i}\,\mathbb{S},\quad
\mathbb{S}_{ij}:=\frac{1}{\lambda_{s}^{2}}\hat D_{i}\hat D_{j}\mathbb{S}+\frac{1}{3}\gamma_{ij}\mathbb{S},\quad
\mathbb{V}_{ij}:=-\frac{1}{\lambda_{v}}\hat D_{(i}\mathbb{V}_{j)},
\label{Si}
\end{equation}
where the non-zero constants $\lambda_{s}$ and $\lambda_{v}$  are roots of eigenvalues of $\mathbb{S}$ and $\mathbb{V}_{i}$, appearing in the equations $(\hat\triangle+\lambda_{s}^{2})\mathbb{S}=0$ and $(\hat\triangle+\lambda_{v}^{2})\mathbb{V}_{i}=0$, respectively.
Note that the tensors $\mathbb{S}_{ij}$ and $\mathbb{V}_{ij}$ are both traceless, i.e. they obey $\mathbb{S}^{i}{}_{i}=0$ and $\mathbb{V}^{i}{}_{i}=0$.
For a given $\mathbb{S}$, $\mathbb{V}_{i}$ or $\mathbb{T}_{ij}$, a mode of the metric perturbation $h_{\lambda\rho}$ is given as either of the following:
\begin{subequations}
\begin{gather}
h_{ab}=f_{ab}(y)\mathbb{S},\quad 
h_{ai}=rf_{a}^{s}(y)\mathbb{S}_{i},\quad 
h_{ij}=2r^{2}\big\{H_{L}(y)\gamma_{ij}\mathbb{S}+H_{T}^{s}(y)\mathbb{S}_{ij}\big\},
\\
h_{ai}=rf_{a}^{v}(y)\mathbb{V}_{i},\quad
h_{ij}=2r^{2}H_{T}^{v}(y)\mathbb{V}_{ij},
\\
h_{ij}=2r^{2}H_{T}^{t}(y)\mathbb{T}_{ij},
\end{gather}
\end{subequations}
whence we have
\begin{subequations}
\begin{gather}
h_{ab}=f_{ab}\mathbb{S},
\quad
h_{a}=-\frac{r}{\lambda_{s}}f_{a}^{s}\,\mathbb{S},\quad
h_{L}=2r^{2}H_{L}\mathbb{S},\quad
h_{T}=\frac{2r^{2}}{\lambda_{s}^{2}}\,H_{T}^{s}\mathbb{S},
\\
h_{T}\,_{ai}=rf_{a}^{v}\mathbb{V}_{i},\quad 
h_{T}\,{}_{i}=-\frac{r^{2}}{\lambda_{v}}H_{T}^{v}\mathbb{V}_{i},
\label{h:harmonicsV}
\\
h_{T}\,{}_{ij}=2r^{2}H_{T}^{t}\mathbb{T}_{ij}.
\label{h:harmonicsT}
\end{gather}
\end{subequations}
Note that, in the investigation of stability analysis of the spacetime, the master variable is constructed from $(f_{ab},f_{a}^{s},H_{L},H_{T}^{s})$, $(f_{a}^{v},H_{T}^{v})$ or $H_{T}^{t}$ for the scalar, vector or tensor harmonics, respectively.
However, we will not mention the master equation in the current investigation of the symmetry operator, that is, we will not impose 
any field equation.

The image $Z_{\mu\nu}$ of the map of 
a single mode of a vector harmonics $\mathbb{V}_{i}$ or a tensor harmonics $\mathbb{T}_{ij}$ by the symmetry operator 
is obtained by substituting Eqs.~\eqref{h:harmonicsV} and \eqref{h:harmonicsT} into Eq.~\eqref{sch:rel3},
\begin{equation}
Z_{T}\,_{ai}=rf_{a}^{v}\,\hat\varepsilon_{i}{}^{kl}\hat D_{k}\mathbb{V}_{l},\quad
Z_{T}\,{}_{j}=-\frac{1}{\lambda_{v}^{2}}r^{2}H_{T}^{v}\,\hat\varepsilon_{j}{}^{kl}\hat D_{k}\mathbb{V}_{l},\quad
Z_{T}\,{}_{ij}=2r^{2}H_{T}^{t}\cdot2\hat\varepsilon_{(i}{}^{kl}\hat D_{|k|}\mathbb{T}_{j)l},
\label{Z:harmonics}
\end{equation}
and we also recall that the scalar harmonics $\mathbb{S}$ has no room to contribute in the map by the symmetry operator.
Therefore, the symmetry operator in the Schwarzschild case is the operation $\hat\varepsilon_{i}{}^{kl}\hat D_{k}$ 
acting on the vector and tensor harmonics as in Eq.~\eqref{Z:harmonics}.

It is known that the vector harmonics $\mathbb{V}_{i}$ and the tensor harmonics $\mathbb{T}_{ij}$ on $S^{3}$ are determined by the scalar harmonics and their derivatives~\cite{doi:10.1063/1.523649,Lindblom:2017maa}.
In the hyperspherical coordinates $(\chi,\vartheta,\varphi)$, we consider a scalar harmonics
\begin{equation}
\mathbb{S}^{\underline{k\ell m}}
=H^{\underline{k\ell}}(\chi)Y^{\underline{\ell m}}(\vartheta,\varphi),
\end{equation}
where $\underline{k}$, $\underline{\ell}$ and $\underline{m}$ are integers characterizing the harmonics, satisfying $\underline{k},\underline{\ell}\ge0$ and $\underline{k}\ge\underline{\ell}\ge|\underline{m}|$, the functions $H^{\underline{k\ell}}$ and $Y^{\underline{\ell m}}$ are decomposed into $H^{\underline{k\ell}}=\sin^{\underline{\ell}}\chi\, C^{\underline{k\ell}}(\chi)$ and $Y^{\underline{\ell m}}=e^{-i\underline{m}\varphi}\Theta^{\underline{\ell m}}(\vartheta)$, and the functions $C^{\underline{k\ell}}$ and $\Theta^{\underline{\ell m}}$ are the solutions of the following differential equations,
\begin{align}
\left[(1-\cos^{2}\chi)\frac{d^{2}}{d(\cos\chi)^{2}}-(2\underline{\ell}+3)\cos\chi\frac{d}{d(\cos\chi)}+\underline{k}(\underline{k}+2)-\underline{\ell}(\underline{\ell}+2)\right]C^{\underline{k\ell}}&=0,
\\
\left[\sin\vartheta\frac{d}{d\vartheta}\left(\sin\vartheta\frac{d}{d\vartheta}\right)+\underline{\ell}(\underline{\ell}+1)\sin^2\vartheta-\underline{m}^{2}\right]\Theta^{\underline{\ell m}}&=0.
\end{align}
For convenience, the superscripts $\underline{k}$, $\underline{\ell}$ and $\underline{m}$ of the spherical harmonics will be  omitted again as long as unnecessary.
The three classes of vector harmonics are given as
\begin{subequations}
\begin{align}
\mathbb{V}^{(0)}_{i}
&:=\frac{1}{\sqrt{\underline{k}(\underline{k}+2)}}
\hat D_{i}\mathbb{S},
\label{V0}
\\
\mathbb{V}^{(1)}_{i}
&:=\frac{1}{\sqrt{\underline{\ell}(\underline{\ell}+1)}}\,
\hat\varepsilon_{i}{}^{jk}\hat D_{j}\mathbb{S}~\hat D_{k}\cos\chi,
\label{V1}
\\
\mathbb{V}^{(2)}_{i}
&:=\frac{1}{\underline{k}+1}\,
\hat\varepsilon_{i}{}^{jk}\hat D_{j}\mathbb{V}^{(1)}_{k},
\label{V2}
\end{align}
\end{subequations}
and the six classes of tensor harmonics are given as
\begin{subequations}
\begin{align}
\mathbb{T}^{(0)}_{ij}
&:=\frac{1}{\sqrt{3}}\,\mathbb{S}\,\gamma_{ij},
\label{T0}
\\
\mathbb{T}^{(1)}_{ij}
&:=\frac{2}{\sqrt{2(\underline{k}-1)(\underline{k}+3)}}
\,\hat D_{(i}\mathbb{V}^{(1)}_{j)},
\\
\mathbb{T}^{(2)}_{ij}
&:=\frac{2}{\sqrt{2(\underline{k}-1)(\underline{k}+3)}}\,\hat D_{(i}\mathbb{V}^{(2)}_{j)},
\\
\mathbb{T}^{(3)}_{ij}
&:=\frac{\sqrt{3}}{\sqrt{2(\underline{k}-1)(\underline{k}+3)}}
\left(\hat D_{i}\mathbb{V}^{(0)}_{j}+\frac{\sqrt{\underline{k}(\underline{k}+2)}}{\sqrt{3}}\mathbb{T}^{(0)}_{ij}\right),
\\
\mathbb{T}^{(4)}_{ij}
&:=
\sqrt{\frac{(\underline{\ell}-1)(\underline{\ell}+2)}{2\underline{k}(\underline{k}+2)}}
\Big\{
E^{\underline{k\ell}}\hat D_{(i}F^{\underline{\ell m}}_{j)}+2\csc^{2}\chi\Big[\frac{\underline{\ell}-1}{2}\cos\chi E^{\underline{k\ell}}+C^{\underline{k\ell}}\Big]F^{\underline{k\ell}}_{(i}\hat D_{j)}\cos\chi\Big\},
\\
\mathbb{T}^{(5)}_{ij}
&:=
\frac{2}{2(\underline{k}+1)}\hat\varepsilon_{(i}{}^{kl}\hat D_{|k|}\mathbb{T}^{(4)}_{j)l},
\label{T5}
\end{align}
\end{subequations}
where
\begin{equation}
E^{\underline{k\ell}}:=-\frac{2\csc^{\underline{\ell}+1}\chi}{(\underline{\ell}-1)(\underline{\ell}+2)}\frac{d}{d\chi}(\sin^{2}\chi H^{\underline{k\ell}}),\quad
F^{\underline{k\ell}}_{a}:=\frac{1}{\sqrt{\underline{\ell}(\underline{\ell}+1)}}\,\hat\varepsilon_{a}{}^{bc}\hat D_{b}(\sin^{\underline{\ell}}\chi Y^{\underline{\ell m}})\hat D_{c}\cos\chi.
\end{equation}
The vectors $\mathbb{S}_{i}$ and the tensors $\mathbb{S}_{ij}$ and $\mathbb{V}_{ij}$, presented in Eq.~\eqref{Si}, are given as some combination of these classes of vector and tensor harmonics.
Also note that these classes of the vector harmonics or the tensor harmonics with every $\underline{k}$, $\underline{\ell}$ and $\underline{m}$ are orthonormal in the following sense:
\begin{equation}
\int\omega_{\gamma}\mathbb{V}^{(A)\underline{k\ell m}}\cdot\mathbb{V}^{(B)\underline{k^{\prime}\ell^{\prime} m^{\prime}}}
=
\int\omega_{\gamma}\mathbb{T}^{(A)\underline{k\ell m}}:\mathbb{T}^{(B)\underline{k'\ell' m'}}=\delta^{AB}\delta^{\underline{kk'}}\delta^{\underline{\ell\ell'}}\delta^{\underline{mm'}},
\end{equation}
where $A$ and $B$ denote the labels for the classes of the vector and tensor harmonics, $\omega_{\gamma}:=\frac{1}{3!}\hat\varepsilon_{ijk}dz^{i}\wedge dz^{j}\wedge dz^{k}$ denotes the volume form of $(S^{3},\gamma)$, and we defined the dot product $\mathbb{V}\cdot\mathbb{V}=\mathbb{V}^{i}\mathbb{V}_{i}$ and double dot product $\mathbb{T}:\mathbb{T}=\mathbb{T}^{ij}\mathbb{T}_{ji}$ for ease of notation.
Also note that Eqs.~\eqref{V0}--\eqref{T5} are eigenvectors and eigentensors of the connection Laplacian $\hat\triangle=\hat D^{i}\hat D_{i}$; readers may refer to Appendix~\ref{app:C1}, where the eigenvalue equation of $\hat\triangle$ and the divergence for the vector and tensor harmonics~\eqref{V0}--\eqref{T5} are exhibited.

We have shown in Eq.~\eqref{Z:harmonics} that the symmetry operator maps a vector harmonics $\mathbb{V}_{i}$ to $\hat\varepsilon_{i}{}^{kl}\hat D_{k}\mathbb{V}_{l}$, and it also maps a tensor harmonics $\mathbb{T}_{ij}$ to $2\hat\varepsilon_{(i}{}^{kl}\hat D_{|k|}\mathbb{T}_{j)l}$.
We uniformly denote the maps of vectors and tensors induced from the symmetry operator as $\hat{\cal S}$.
Comparing $\hat{\cal S}$ with the definitions~\eqref{V0}--\eqref{T5} of the classes of $\mathbb{V}_{i}$ and $\mathbb{T}_{ij}$, we have that the map by the symmetry operator results in
\begin{subequations}
\begin{gather}
\hat{\cal S}\,\mathbb{V}^{(0)}=0,
\quad
\hat{\cal S}\begin{bmatrix}\mathbb{V}^{(1)}\\\mathbb{V}^{(2)}\end{bmatrix}=(\underline{k}+1)\begin{bmatrix}\mathbb{V}^{(2)}\\\mathbb{V}^{(1)}\end{bmatrix},
\\
\hat{\cal S}\,\mathbb{T}^{(0)}=0,
\quad
\hat{\cal S}\begin{bmatrix}\mathbb{T}^{(1)}\\\mathbb{T}^{(2)}\end{bmatrix}=(\underline{k}+1)\begin{bmatrix}\mathbb{T}^{(2)}\\\mathbb{T}^{(1)}\end{bmatrix},
\quad
\hat{\cal S}\,\mathbb{T}^{(3)}= 0,
\quad
\hat{\cal S}\begin{bmatrix}\mathbb{T}^{(4)}\\\mathbb{T}^{(5)}\end{bmatrix}=2(\underline{k}+1)\begin{bmatrix}\mathbb{T}^{(5)}\\\mathbb{T}^{(4)}\end{bmatrix}.
\end{gather}
\end{subequations}
To summarize, in the Schwarzschild case, the symmetry operator reduces to the operation $\hat{\cal S}$ on $S^{3}$ that maps a class of vector or tensor harmonics to zero, or otherwise $\hat{\cal S}$ maps a class of harmonics to the paired class.
In other words, we have found the eigenvectors and eigentensors of $\hat{\cal S}$; in addition to $\mathbb{V}^{(0)}$, $\mathbb{T}^{(0)}$ and $\mathbb{T}^{(3)}$, having zero eigenvalue, $\hat{\cal S}$ maps the combinations $\mathbb{V}^{(1)}\pm\mathbb{V}^{(2)}$, $\mathbb{T}^{(1)}\pm\mathbb{T}^{(2)}$ and $\mathbb{T}^{(4)}\pm\mathbb{T}^{(5)}$ to themselves except for multiplication by constant.
It is worth noting that $\mathbb{V}^{(1)}\pm\mathbb{V}^{(2)}$ with $\underline{k}=1$ correspond to the six independent Killing vectors on $(S^{3},\gamma)$; see Appendix~\ref{app:C1}.

\section{Finite angular momenta case}
\label{sec:III}

\subsection{Metric and Killing-Yano 3-from in designated frame}
\label{sec:4A}

While the metric, the orthonormal frame and the Killing-Yano 3-form in the finite momenta case have been given in Eqs.~\eqref{met:equal}--\eqref{KY3form:equal} in Sec.~\ref{sec:IIB}, where the Boyer-Lindquist coordinates are adopted, we use another frame compatible with $SU(2)\times SU(2)$ isometry of $(S^{3},\gamma)$, following Refs.~\cite{Murata:2007gv,Murata:2008yx}.
Via the coordinate transformation $\tilde\theta=2\theta$, $\tilde\phi=\phi+\psi$ and $\tilde\psi=-\phi+\psi$ of the angular coordinate system,
we introduce the covector fields $\sigma^{i}$ with $i\in\{1,2,3\}$ by
\begin{subequations}
\begin{align}
\sigma^{1}&:=-\sin\tilde\phi~d\tilde\theta+\cos\tilde\phi\sin\tilde\theta~d\tilde\psi,\label{s1}\\
\sigma^{2}&:=-\cos\tilde\phi~d\tilde\theta-\sin\tilde\phi\sin\tilde\theta~d\tilde\psi,\label{s2}\\
\sigma^{3}&:=d\tilde\phi+\cos\tilde\theta~d\tilde\psi
. \label{s3}
\end{align}
\end{subequations}
Note that $\sigma^{i}$ are the dual of the generators $\sigma_{i}$ of $SO(3)$ isometry of $(S^{3},\gamma)$, as we will remark in the next subsection, and hence $\sigma^{i}$ obey $d\sigma^{i}=-\frac{1}{2}\epsilon^{i}{}_{jk}\sigma^{j}\wedge \sigma^{k}$, 
where $\epsilon^{i}{}_{jk}$ denotes the Levi-Civita symbol.
In practice, we use alternative covector fields $\bar\sigma^{i}:=\frac{1}{2}\sigma^{i}$ for convenience, where $\bar\sigma^{i}$ obey $d\bar\sigma^{i}=-\epsilon^{i}{}_{jk}\bar\sigma^{j}\wedge \bar\sigma^{k}$.
We work on the finite angular momenta case in the frame $(dt,d\tilde r,\bar\sigma^{+},\bar\sigma^{-},\bar\sigma^{3})$, where $\bar\sigma^{\pm}:=\frac{1}{\sqrt{2}}(\bar\sigma^{1}\pm {\rm i}\bar\sigma^{2})$ are the null bases with ${\rm i}=\sqrt{-1}$ being the imaginary unit, and $\tilde r$ is the alternative radial coordinate defined by $\tilde r^{2}:=\Sigma=r^{2}+a^{2}$.
Eqs.~\eqref{met:equal}--\eqref{KY3form:equal} are rewritten as
\begin{equation}
g=-dt^{2}+\frac{d\tilde r^{2}}{G(\tilde r)}+\tilde r^{2}\Big\{(\bar\sigma^{+}\bar\sigma^{-}+\bar\sigma^{-}\bar\sigma^{+})+(\bar\sigma^{3})^{2}\Big\}+\frac{\mu}{\tilde r^{2}}\left(dt-a\bar\sigma^{3}\right)^{2},
\label{met:frame}
\end{equation}
\begin{subequations}
\begin{align}
e^{0}&=\sqrt{\frac{\tilde r^{2}G(\tilde r)}{\tilde r^{2}-a^{2}}}~\Big(dt -a\,\bar\sigma^{3}\Big),\\
e^{1}&=\frac{\tilde r}{\sqrt{\tilde r^{2}G(\tilde r)}}~d\tilde r,\\
e^{2}&=-\frac{{\rm i}\tilde r}{\sqrt{2}}\left(-{\rm e}^{{\rm i}\tilde \phi}\,\bar\sigma^{+}+{\rm e}^{-{\rm i}\tilde\phi}\,\bar\sigma^{-}\right),\\
e^{3}&=-\frac{\tilde r}{\sqrt{2}}\left({\rm e}^{{\rm i}\tilde\phi}\,\bar\sigma^{+}+{\rm e}^{-{\rm i}\tilde\phi}\,\bar\sigma^{-}\right),\\
e^{5}&=\frac{1}{\sqrt{\tilde r^{2}-a^{2}}}\Big(-adt +\tilde r^{2}\,\bar\sigma^{3}\Big),
\end{align}
\end{subequations}
and 
\begin{equation}
f=-a\tilde r\,dt\wedge d\tilde r\wedge \bar\sigma^{3}-{\rm i}\tilde r^{2}\,\bar\sigma^{+}\wedge \bar\sigma^{-}\wedge \left(-a\,dt+\tilde r^{2}\,\bar\sigma^{3}\right),
\label{equal:KY}
\end{equation}
where
\begin{equation}
G(\tilde r):=1-\frac{\mu}{\tilde r^{2}}+\frac{\mu a^{2}}{\tilde r^{4}},\qquad \mu:=2M.
\end{equation}

In preparation for the later calculation, let us summarize the non-zero components of the metric, the Killing-Yano 3-form, the inverse metric and the connection coefficients.
The non-zero components of the metric~\eqref{met:frame} are 
\begin{equation}
g_{tt}=-1+\frac{\mu}{\tilde r^{2}},\quad g_{t3}=-\frac{\mu a}{\tilde r^{2}},\quad g_{33}=\tilde r^{2}+\frac{\mu a^{2}}{\tilde r^{2}},\quad
g_{\tilde r\tilde r}=G(\tilde r)^{-1},\quad g_{+-}=\tilde r^{2}.
\end{equation}
Note that we have $g_{tt}g_{33}-g_{t3}^{2}=-\tilde r^{2}G(\tilde r)$, and then $\det g=\tilde r^{6}$.
The non-zero components of the Killing-Yano 3-form~\eqref{equal:KY} are 
\begin{equation}
f_{t\tilde r 3}=-a\tilde r,\quad f_{+-t}={\rm i}a\tilde r^{2},\quad f_{+- 3}=-{\rm i}\tilde r^{4}.
\end{equation}
Then we present the inverse metric and the connection coefficients.
We define the vector fields 
$\bar\sigma_{i}=2\sigma_{i}$, the dual of $\bar\sigma^{i}$, 
satisfying the relations $\bar\sigma_{i}\cdot\bar\sigma^{j}=\delta_{i}{}^{j}$ and $\bar\sigma_{i}\cdot dt=\bar\sigma_{i}\cdot d\tilde r=0$, and we also define the null bases $\bar\sigma_{\pm}:=\frac{1}{\sqrt{2}}(\bar\sigma_{1}\mp {\rm i}\bar\sigma_{2})$, satisfying the relations $\bar\sigma_{+}\cdot\bar\sigma^{+}=\bar\sigma_{-}\cdot\bar\sigma^{-}=1$ and $\bar\sigma_{+}\cdot\bar\sigma^{-}=0$ with $\bar\sigma^{\pm}$.
In the frame $(\partial_{t},\partial_{r},\bar\sigma_{+},\bar\sigma_{-},\bar\sigma_{3})$, the inverse metric is given by
\begin{equation}
g^{-1}=-(\partial_{t})^{2}+G(\tilde r)(\partial_{\tilde r})^{2}+\frac{1}{\tilde r^{2}}\Big\{(\bar\sigma_{+}\bar\sigma_{-}+\bar\sigma_{-}\bar\sigma_{+})+(\bar\sigma_{3})^{2}\Big\}-\frac{\mu}{\tilde r^{2}G(\tilde r)}\left(\partial_{t}-\frac{a}{\tilde r^{2}}\bar\sigma_{3}\right)^{2}.
\end{equation}
Therefore, the non-zero components of $g^{-1}$ are 
\begin{equation}
g^{tt}=-1-\frac{\mu}{\tilde r^{2}G(\tilde r)},\quad g^{t3}=-\frac{\mu a}{\tilde r^{4}G(\tilde r)},\quad
g^{33}=\frac{1}{\tilde r^{2}}\left(1-\frac{\mu a^{2}}{\tilde r^{4}G(\tilde r)}\right),\quad g^{rr}=G(\tilde r),\quad g^{+-}=\frac{1}{\tilde r^{2}}.
\end{equation}
Let $\bar\sigma^{\mu}$ represent a basis of the covector frame $(dt,d\tilde r,\bar\sigma^{+},\bar\sigma^{-},\bar\sigma^{3})$, and let $\bar\sigma_{\mu}$ represent a basis of the vector frame $(\partial_{t},\partial_{r},\bar\sigma_{+},\bar\sigma_{-},\bar\sigma_{3})$.
The connection coefficients, defined by 
$\Gamma^{\mu}{}_{\nu\rho}:=\bar\sigma^{\mu}\cdot\nabla_{\bar\sigma_{\nu}}\bar\sigma_{\rho}$,
are calculated as follows:
\begin{gather}
\Gamma^{t}{}_{t\tilde r}=\Gamma^{t}{}_{\tilde r t}=\frac{\mu}{\tilde r^{3}G(\tilde r)},\quad
\Gamma^{t}{}_{\tilde r 3}=\Gamma^{t}{}_{3\tilde r}=-\frac{2\mu a}{\tilde r^{3}G(\tilde r)},\quad
\Gamma^{\tilde r}{}_{tt}=\frac{\mu G(\tilde r)}{\tilde r^{3}},\quad
\Gamma^{\tilde r}{}_{t 3}=\Gamma^{\tilde r}{}_{3 t}=-\frac{\mu a G(\tilde r)}{\tilde r^{3}},
\notag\\
\Gamma^{\tilde r}{}_{\tilde r\tilde r}=\frac{1}{\tilde r}-\left(1-\frac{\mu a^{2}}{\tilde r^{4}}\right)\frac{1}{\tilde r G(\tilde r)},\quad
\Gamma^{\tilde r}{}_{+-}=\Gamma^{\tilde r}{}_{-+}=-\tilde r G(\tilde r),\quad
\Gamma^{\tilde r}{}_{33}=-\left(1-\frac{\mu a^{2}}{\tilde r^{4}}\right)\tilde r G(\tilde r),
\notag\\
\Gamma^{+}{}_{t+}=\Gamma^{+}{}_{+t}=-\Gamma^{-}{}_{t-}=-\Gamma^{-}{}_{-t}=\frac{{\rm i}\mu a}{\tilde r^{4}},\quad
\Gamma^{+}{}_{\tilde r +}=\Gamma^{+}{}_{+\tilde r}=\Gamma^{-}{}_{\tilde r-}=\Gamma^{-}{}_{-\tilde r}=\frac{1}{\tilde r},
\notag\\
\Gamma^{+}{}_{+ 3}=-\Gamma^{-}{}_{- 3}=-{\rm i}\left(1+\frac{\mu a^{2}}{\tilde r^{4}}\right),\quad
\Gamma^{+}{}_{3 +}=-\Gamma^{-}{}_{3-}={\rm i}\left(1-\frac{\mu a^{2}}{\tilde r^{4}}\right),
\notag\\
\Gamma^{3}{}_{t\tilde r}=\Gamma^{3}{}_{\tilde r t}=\frac{\mu a}{\tilde r^{5}G(\tilde r)},\quad
\Gamma^{3}{}_{\tilde r 3}=\Gamma^{3}{}_{3 \tilde r}=\frac{1}{\tilde r}\left(1-\frac{2\mu a^{2}}{\tilde r^{4}G(\tilde r)}\right),\quad
\Gamma^{3}{}_{+-}=-\Gamma^{3}{}_{-+}={\rm i}.
\label{ConnectionCoeff}
\end{gather}

\subsection{Mode decomposition of rank-2 symmetric tensors}
\label{sec:4B}

In the analysis in the finite angular momenta case, let us give the mode decomposition of linear metric perturbations of the spacetime from the perspective of the $\mathfrak{su}(2)\times \mathfrak{su}(2)$ algebra for $S^{3}$, following Refs.~\cite{Murata:2007gv,Murata:2008yx}.
As is well known, the metric $\gamma$ of the unit 3-sphere $S^{3}$ in the Hopf coordinates $(\theta,\phi,\psi)$ is given by $\gamma=d\theta^{2}+\sin^{2}\theta d\phi^{2}+\cos^{2}\theta d\psi^{2}$, and, in the current frame, 
\begin{equation}
\gamma
=(\bar\sigma^{1})^{2}+(\bar\sigma^{2})^{2}+(\bar\sigma^{3})^{2}.
\label{gamma:sigma}
\end{equation}
The volume form $\omega_{g}$ of the spacetime is obtained as
\begin{equation}
\omega_{g}
=\tilde r^{3}dt\wedge d\tilde r\wedge \omega_{\gamma},
\label{vol}
\end{equation}
where 
$\omega_{\gamma}:=\bar\sigma^{1}\wedge\bar\sigma^{2}\wedge\bar\sigma^{3}$ is the volume form of $(S^{3},\gamma)$.
This implies that the orthonormal bases in the square-integrable function space on $S^{3}$ can be applied to the construction of the modes of rank-2 symmetric tensors such as linear metric perturbations.
We begin the review of the characterization of the orthonormal bases in the function space, referred to as the Wigner $\mathbb{D}$ functions in Refs.~\cite{Murata:2007gv,Murata:2008yx}, with introducing the following six independent Killing vector fields on $(S^{3},\gamma)$:
\begin{subequations}
\begin{align}
\sigma_{1}&=-\sin\tilde\phi\,\partial_{\tilde\theta}
-\cot\tilde\theta\cos\tilde\phi~\partial_{\tilde\phi}
+\csc\tilde\theta\cos\tilde\phi\,\partial_{\tilde\psi},
\label{s1vec}\\
\sigma_{2}&=-\cos\tilde\phi~\partial_{\tilde\theta}
+\cot\tilde\theta\sin\tilde\phi~\partial_{\tilde\phi}
-\csc\tilde\theta\sin\tilde\phi\,\partial_{\tilde\psi},\\
\sigma_{3}&=\partial_{\tilde\phi},\label{s3vec}
\end{align}
and
\begin{align}
\xi_{1}&=-\sin\tilde\psi\,\partial_{\tilde\theta}+\csc\tilde\theta\cos\tilde\psi\,\partial_{\tilde\phi}-\cot\tilde\theta\cos\tilde\psi\,\partial_{\tilde\psi},
\\
\xi_{2}&=-\cos\tilde\psi\,\partial_{\tilde\theta}
-\csc\tilde\theta\sin\tilde\psi\,\partial_{\tilde\phi}
+\cot\tilde\theta\sin\tilde\psi\,\partial_{\tilde\psi},
\\
\xi_{3}&=\partial_{\tilde\psi},
\label{xi3vec}
\end{align}
\end{subequations}
where we recall that $\sigma_{i}$ are the dual vectors of $\sigma^{i}$.
Note that $\sigma_{1}$ and $\sigma_{2}$ are not Killing vector fields on the spacetime while $\sigma_{3}$ and $\xi_{i}$ are the Killing vectors on the spacetime.
The Lie algebra $\mathfrak{su}(2)\times\mathfrak{su}(2)$ is spanned by $W_{i}:={\rm i}\sigma_{i}$ and $L_{i}:={\rm i}\xi_{i}$, obeying
\begin{equation}
[W_{i},W_{j}]={\rm i}\epsilon^{k}{}_{ij}W_{k},\quad [L_{i},L_{j}]={\rm i}\epsilon^{k}{}_{ij}L_{k},\quad[W_{i},L_{j}]=0.
\end{equation}
We also have the Casimir operator $W^{2}:=\sum_{i}W_{i}^{2}=\sum_{i}L_{i}^{2}=:L^{2}$ and the ladder operators $W_{\pm}:=W_{1}\pm {\rm i}W_{2}$ and $L_{\pm}:=L_{1}\pm {\rm i}L_{2}$.
The Wigner $\mathbb{D}$ functions 
are complex-valued eigenfunctions of $W_{3}$, $L_{3}$ and $W^{2}=L^{2}$, labeled by the labels $(J,K,M)$, obeying
\footnote{
The allowed labels $(J,K,M)$ of the Wigner $\mathbb{D}$ functions satisfy the following three conditions: 
\begin{enumerate}
\item
$(J,K,M)$ satisfy the inequalities $J\ge0$ and $2J\ge|K|+|M|$,
\item
$(J,K,M)$ are all integers, or $(J,K,M)$ are all half-integers, 
\item
$|K|+|M|$ is even for an integer $J$, or $|K|+|M|$ is odd for a half-integer $J$.
\end{enumerate}
}
\begin{gather}
W_{3}\mathbb{D}^{J}_{KM}=K\mathbb{D}^{J}_{KM},\quad L_{3}\mathbb{D}^{J}_{KM}=M\mathbb{D}^{J}_{KM},\quad W^{2}\mathbb{D}^{J}_{KM}=L^{2}\mathbb{D}^{J}_{KM}=J(J+1)\mathbb{D}^{J}_{KM},
\\
W_{\pm}\mathbb{D}^{J}_{KM}=\sqrt{J(J+1)-K(K\pm1)}\,\mathbb{D}^{J}_{K\pm1\,M},\quad
L_{\pm}\mathbb{D}^{J}_{KM}=\sqrt{J(J+1)-M(M\pm1)}\,\mathbb{D}^{J}_{K\,M\pm1}.\notag
\end{gather}
We also remark that $W_{i}$ and $L_{i}$ also act on the bases $\bar\sigma^{3}$ and $\bar\sigma^{\pm}$.
From the coordinate representations~\eqref{s1}--\eqref{s3} of $\sigma^{i}$ and the coordinate representation~\eqref{s1vec}--\eqref{xi3vec} of the Killing vectors $\sigma_{i}$ and $\xi_{i}$, we have
\begin{equation}
\pounds_{\sigma_{i}}\sigma^{j}=\epsilon_{ki}{}^{j}\sigma^{k},\quad
\pounds_{\xi_{i}}\sigma^{j}=0,
\label{W:sigma}
\end{equation}
where $\pounds$ denotes the Lie derivative, and $\epsilon_{ki}{}^{j}$ denotes the Levi-Civita symbol, whence
\begin{gather}
\pounds_{W_{3}}\bar\sigma^{\pm}=\pm\bar\sigma^{\pm},\quad
\pounds_{W_{\mp}}\bar\sigma^{\pm}=\mp\sqrt{2}\bar\sigma^{3},\quad
\pounds_{W_{\mp}}\bar\sigma^{3}=\pm\sqrt{2}\bar\sigma^{\mp},\quad
\pounds_{W_{3}}\bar\sigma^{3}=\pounds_{W_{\pm}}\bar\sigma^{\pm}=0,
\notag\\
\pounds_{L_{3}}\bar\sigma^{3}=\pounds_{L_{3}}\bar\sigma^{\pm}=\pounds_{L_{\pm}}\bar\sigma^{\pm}=\pounds_{L_{\mp}}\bar\sigma^{\pm}=0.
\label{W:sigmapm}
\end{gather}

The mode decomposition of a rank-2 symmetric tensor has the labels $(J,K,M)$ as well as the Wigner $\mathbb{D}$ functions.
For convenience, we denote the covector bases as $\bar\sigma^{t}:=dt$ and $\bar\sigma^{t}:=d\tilde r$ together with $\bar\sigma^{3}$ and $\bar\sigma^{\pm}$, and introduce the indices $A,B,\cdots\in\{t,\tilde r,3\}$.
We can expand the tensor $h$ as
$h=\sum_{J,K,M}h^{J}_{KM}$,
where each mode $h^{J}_{KM}$ is defined as
\begin{align}
h^{J}_{KM}&:=h_{AB}^{JKM}(t,\tilde r)\mathbb{D}^{J}_{KM}\bar\sigma^{A}\bar\sigma^{B}
+2h_{+-}^{JKM}(t,\tilde r)\mathbb{D}^{J}_{KM}\bar\sigma^{+}\bar\sigma^{-}
\notag\\
&\quad
+2h_{A+}^{JKM}(t,\tilde r)\mathbb{D}^{J}_{K-1\,M}\bar\sigma^{A}\bar\sigma^{+}
+2h_{A-}^{JKM}(t,\tilde r)\mathbb{D}^{J}_{K+1\,M}\bar\sigma^{A}\bar\sigma^{-}
\notag\\
&\quad
+h_{++}^{JKM}(t,\tilde r)\mathbb{D}^{J}_{K-2\,M}\bar\sigma^{+}\bar\sigma^{+}
+h_{--}^{JKM}(t,\tilde r)\mathbb{D}^{J}_{K+2\,M}\bar\sigma^{-}\bar\sigma^{-}.
\label{hJKM}
\end{align}
Note that the complex conjugate of $\mathbb{D}^{J}_{KM}$ and $h$ are given as $(\mathbb{D}^{J}_{KM})^{*}=\mathbb{D}^{J}_{-K\,-M}$ and $h^{*}=h$, respectively, and therefore the complex conjugate of $h^{J}_{KM}$ is given as $(h^{J}_{KM})^{*}=h^{J}_{-K\,-M}$.
Taking the action \eqref{W:sigmapm} of $W_{3}$ and $L_{3}$ on $\bar\sigma^{\pm}$ into account, 
we can find 
\begin{equation}
\pounds_{W_{3}}h^{J}_{KM}=Kh^{J}_{KM},\quad\pounds_{L_{3}}h^{J}_{KM}=Mh^{J}_{KM}, 
\end{equation}
that is, $h^{J}_{KM}$ is an eigentensor of $W_{3}$ and $L_{3}$.

\subsection{Investigation of the symmetry operator}

We then move on to the investigation of the map of each modes of the tensor $h_{\lambda\rho}$
by the symmetry operator.
For convenience, let us introduce the notation that the basis vectors are denoted as 
$(\bar\sigma_{A},\bar\sigma_{\pm})$, where $A\in\{t,r,3\}$,
as well as the covector bases.
The image $Z_{\lambda\rho}={\cal S}h_{\lambda\rho}$,
given in Eq.~\eqref{solN}, takes the following form:
\begin{subequations}
\begin{align}
Z_{AB}&=
2f_{(A|}{}^{CD}\Big[\bar\sigma_{C}h_{|B)D}-\Gamma^{E}{}_{C|B)}h_{ED}-\Gamma^{E}{}_{CD}h_{|B)E}\Big]
\label{equal:ZAB}
\notag\\
&\quad
+2f_{(A|}{}^{+-}\Big[\bar\sigma_{+}h_{|B)-}-\bar\sigma_{-}h_{|B)+}-(\Gamma^{+}{}_{+|B)}-\Gamma^{-}{}_{-|B)})h_{+-}-(\Gamma^{C}{}_{+-}-\Gamma^{C}{}_{-+})h_{|B)C}\Big],
\\
Z_{+-}&=f_{+}{}^{+A}\Big[\bar\sigma_{+}h_{-A}-\bar\sigma_{-}h_{+A}-(\Gamma^{B}{}_{+-}-\Gamma^{B}{}_{-+})h_{BA}-(\Gamma^{+}{}_{+A}-\Gamma^{-}{}_{-A})h_{+-}\Big],
\label{equal:Z+-}
\\
Z_{A\pm}&=f_{A}{}^{BC}\Big[\bar\sigma_{B}h_{\pm C}-\Gamma^{\pm}{}_{B\pm}h_{\pm C}-\Gamma^{D}{}_{BC}h_{\pm D}\Big]\notag\\
&\quad
+f_{A}{}^{+-}\Big[\bar\sigma_{+}h_{\pm -}-\bar\sigma_{-}h_{\pm +}
-(\mp\Gamma^{B}{}_{\mp\pm}+\Gamma^{B}{}_{+-}-\Gamma^{B}{}_{-+})h_{B \pm}\Big]\notag\\
&\quad
+f_{\pm}{}^{\pm B}\Big[\bar\sigma_{\pm}h_{AB}-\bar\sigma_{B}h_{A\pm}+\Gamma^{C}{}_{BA}h_{C\pm}-\Gamma^{\pm}{}_{\pm A}h_{\pm B}-(\Gamma^{\pm}{}_{\pm B}-\Gamma^{\pm}{}_{B \pm})h_{A\pm}\Big],
\\
Z_{\pm\pm}&=2f_{\pm}{}^{\pm A}\Big[\bar\sigma_{\pm}h_{A\pm}-\bar\sigma_{A}h_{\pm\pm}+(2\Gamma^{\pm}{}_{A\pm}-\Gamma^{\pm}{}_{\pm A})h_{\pm\pm}\Big].
\label{equal:Z++}
\end{align}
\end{subequations}
It is worth noting that we can apply the equalities $f_{+}{}^{+A}(\Gamma^{-}{}_{-A}-\Gamma^{+}{}_{+A})=2$ and $f_{+}{}^{+A}(2\Gamma^{+}{}_{A+}-\Gamma^{+}{}_{+ A})=f_{-}{}^{- A}(2\Gamma^{-}{}_{A-}-\Gamma^{-}{}_{-A})=3$ to Eqs.~\eqref{equal:Z+-} and \eqref{equal:Z++}, respectively, in obtaining the more detailed representation of $Z_{\lambda\rho}$ by writing the connection coefficients~\eqref{ConnectionCoeff} explicitly.
Since the operators $W_{3}$ and $W_{\pm}$ are included in Eqs.~\eqref{equal:ZAB}--\eqref{equal:Z++} in the form $\bar\sigma_{\pm}=-\sqrt{2}\,{\rm i}\,W_{\mp}$ and $\bar\sigma_{3}=-2{\rm i}\,W_{3}$, 
substituting a single mode $h=h^{J}_{KM}$, given in Eq.~\eqref{hJKM}, into Eqs.~\eqref{equal:ZAB}--\eqref{equal:Z++}, we obtain 
\begin{equation}
Z_{AB}\propto \mathbb{D}^{J}_{KM},\quad 
Z_{+-}\propto \mathbb{D}^{J}_{KM}\quad 
Z_{A\pm}\propto \mathbb{D}^{J}_{K\mp1\,M}\quad 
Z_{\pm\pm}\propto \mathbb{D}^{J}_{K\mp2\,M}.
\end{equation}
That is, the symmetry operator maps a single mode $(J,K,M)$ to another tensor labeled by the same $(J,K,M)$.

Here, we remark that we have not applied the field equation for $h_{\mu\nu}$, that is, $h_{\mu\nu}$ is not necessarily a linear metric perturbation.
The result can be interpreted in terms of the commutation relations for the operators as follows.
We denote the Lie derivatives of tensors with respect to $W_{i}$ and $L_{i}$ as ${\cal W}_{i}:=\pounds_{W_{i}}$ and ${\cal L}_{i}:=\pounds_{L_{i}}$, respectively.
From the above result, we can confirm that ${\cal S}$ and ${\cal W}_{3}$ commute with each other,
\begin{equation}
{\cal S}{\cal W}_{3}h
={\cal S}{\cal W}_{3}\sum_{JKM}h^{J}_{KM}
=\sum_{JKM}K{\cal S}h^{J}_{KM}
={\cal W}_{3}{\cal S}\sum_{JKM}h^{J}_{KM}
={\cal W}_{3}{\cal S}h,
\end{equation}
and in the similar way we can also confirm that ${\cal S}$ and ${\cal L}_{i}$ commute with each other.
As a result, the commutation relations among the operators $({\cal S},{\cal W}_{3},{\cal L}_{i})$ are summarized as follows:
\begin{equation}
[{\cal S},{\cal W}_{3}]=0,\quad
[{\cal S},{\cal L}_{i}]=0,\quad
[{\cal W}_{3},{\cal L}_{i}]=0,
\end{equation}
originated from the $U(1)\times SU(2)$ isometry and the antisymmetric hidden symmetry of the 5-dimensional Myers-Perry spacetime with equal angular momenta.
Note that the operators $({\cal S},{\cal W}_{3},{\cal L}_{i})$ commute with the operator ${\cal T}:=\pounds_{{\rm i}\bar\sigma_{t}}$ associated with the stationarity of the spacetime.

Hereafter, we impose the field equation ${\cal M}h=0$, that is, we suppose that $h_{\mu\nu}$ is a linear metric perturbation.
From Theorem~\ref{thm1}, the operator ${\cal M}$ commute with $({\cal S},{\cal W}_{3},{\cal L}_{i},{\cal T})$,
\begin{equation}
[{\cal S},{\cal M}]=0,\quad
[{\cal W}_{3},{\cal M}]=0,\quad
[{\cal L}_{i},{\cal M}]=0,\quad
[{\cal T},{\cal M}]=0.
\end{equation}
Note that the separability of linear metric perturbations is guaranteed by $({\cal W}_{3},{\cal L}_{i},{\cal T})$, associated with $U(1)\times SU(2)\times\mathbb{R}$, as is shown in Eq.~\eqref{hJKM}, and the symmetry operator ${\cal S}$ is redundant from the perspective of separability.

Still there is possibility that the symmetry operator ${\cal S}$ maps a linear metric perturbation to another perturbation to which different boundary condition is imposed, or otherwise ${\cal S}$ is some linear combination of $({\cal W}_{3},{\cal L}_{3},{\cal T})$ and the identity operator ${\cal I}$ which maps $h_{\mu\nu}$ to itself.
In the following subsections, we investigate the map of individual modes by ${\cal S}$ in more detail;
in this paper we investigate the three kinds of modes that were dealt with in Ref.~\cite{Murata:2008yx}: $(J,K,M)=(0,0,0)$, $(J,K,M)=(0,1,0)$ and $K=J+2$.
As a result, the maps of linear metric perturbations with the above modes result in the linear combinations of the identity and the phase shift, that is, the boundary condition is preserved by ${\cal S}$.
Note that we also present the Schwarzschild limit of these modes, and compare with the result of Sec.~\ref{sec:II} in Appendix~\ref{app:C}, where the effect of 
the phase shift vanishes in the limit $a\to0$.

\subsection{$(J,K,M)=(0,0,0)$ mode}
\label{sec:K=0}

From Eq.~\eqref{hJKM},  the $(J,K,M)=(0,0,0)$ mode of a rank-2 symmetric tensor takes the form
\begin{equation}
h
=h^{J=0}_{K=0\,M=0}
=h_{AB}(t,\tilde r)\bar\sigma^{A}\bar\sigma^{B}+2h_{+-}(t,\tilde r)\bar\sigma^{+}\bar\sigma^{-},
\label{K=0:h}
\end{equation}
where $A,B\in\{t,\tilde r,3\}$.
The gauge transformation $\delta h_{\mu\nu}=\nabla_{\mu}\xi_{\nu}+\nabla_{\nu}\xi_{\mu}$ for the $(J,K,M)=(0,0,0)$ mode with respect to a gauge field 
$\xi=\xi_{A}(t,\tilde r)\bar\sigma^{A}$
is calculated to be
\begin{subequations}
\begin{align}
\delta h_{tt}&=2\left(\partial_{t}\xi_{t}-\frac{\mu G(\tilde r)}{\tilde r^{3}}\xi_{\tilde r}\right),\\
\delta h_{t3}&=\partial_{t}\xi_{3}+\frac{2\mu a G(\tilde r)}{\tilde r^{3}}\xi_{\tilde r},\\
\delta h_{33}&=2\tilde r G(\tilde r)\left(1-\frac{\mu a^{2}}{\tilde r^{4}}\right)\xi_{\tilde r},\label{K=0:gauge:33}
\\
\delta h_{\tilde r\tilde r}&=2\partial_{\tilde r}\xi_{\tilde r}+\frac{2}{\tilde r G(\tilde r)}\left(\frac{\mu}{\tilde r^{2}}-\frac{2\mu a^{2}}{\tilde r^{4}}\right)\xi_{\tilde r},\\
\delta h_{t\tilde r}&=\partial_{t}\xi_{\tilde r}+\partial_{\tilde r}\xi_{t}-\frac{2\mu}{\tilde r^{3}G(\tilde r)}\xi_{t}-\frac{2\mu a}{\tilde r^{5} G(\tilde r)}\xi_{3},\\
\delta h_{\tilde r3}&=\partial_{\tilde r}\xi_{3}+\frac{4\mu a}{\tilde r^{3}G(\tilde r)}\xi_{t}-\frac{2}{\tilde r G(\tilde r)}\left(1-\frac{\mu}{\tilde r^{2}}-\frac{\mu a^{2}}{\tilde r^{4}}\right)\xi_{3},\\
\delta h_{+-}&=2\tilde r G(\tilde r)\xi_{\tilde r}.
\label{K=0:gauge:+-}
\end{align}
\end{subequations}
Applying this gauge transformation to modes with non-zero frequency $\omega\neq0$, 
\footnote{Ref.~\cite{Murata:2008yx} does not cover the stationary perturbation $\omega=0$%
, which cannot be compatible with 
the gauge fixing~\eqref{K=0:fix}.
We show the results of the investigation of the $\omega=0$ case in Appendix~\ref{app:stationary} independently of the $\omega\neq0$ modes.}
 we set the following three components to zero:
\begin{equation}
h_{tt}=0,\quad h_{t 3}=0,\quad h_{3 3}=0.
\label{K=0:fix}
\end{equation}
So far the remaining components are $h_{\tilde r\tilde r}$, $h_{t\tilde r}$, $h_{\tilde r3}$ and $h_{+-}$.
From Eqs.~\eqref{K=0:gauge:33} and \eqref{K=0:gauge:+-}, we find that one of the gauge invariant variables is given as the combination of $h_{33}$ and $h_{+-}$, and it has been reduced to $h_{+-}$ under the current gauge fixing.
According to Ref.~\cite{Murata:2008yx}, eliminating $h_{\tilde r\tilde r}$, $h_{t\tilde r}$ and $h_{\tilde r3}$ from the perturbation equation for $h_{\lambda\rho}$, we can obtain the master equation for the master variable
\begin{equation}
\Phi_{0}
:=\frac{(\tilde r^{4}-\mu a^{2})(\tilde r^{4}+\mu a^{2})^{1/4}}{\tilde r^{3/2}(3\tilde r^{4}+\mu a^{2})}
h_{+-}.
\end{equation}

Substituting Eq.~\eqref{K=0:h} together with gauge fixing~\eqref{K=0:fix} into Eqs.~\eqref{equal:ZAB}--\eqref{equal:Z+-}, we have $Z_{\lambda\rho}={\cal S}h_{\lambda\rho}$ for the $(J,K,M)=(0,0,0)$ mode as follows:
\begin{subequations}
\begin{align}
Z_{tt}&=2\left\{-\frac{\mu a^{2}}{\tilde r^{3}}\partial_{t}h_{t\tilde r}
+\frac{\mu a^{2}G(\tilde r)}{\tilde r^{4}} h_{\tilde r\tilde r}
-\frac{2\mu a^{2}}{\tilde r^{6}}h_{+-}\right\},
\label{K=0:gaugeN:tt}
\\
Z_{t 3}&=-\frac{\mu a^{2}}{\tilde r^{3}}\partial_{t}h_{\bar 3\tilde r}+a\tilde r\left(1+\frac{\mu a^{2}}{\tilde r^{4}}\right)\partial_{t}h_{t\tilde r}
+aG(\tilde r)\left(1-\frac{2\mu}{\tilde r^{2}}-\frac{\mu a^{2}}{\tilde r^{4}}\right)h_{\tilde r\tilde r}
+\frac{2a}{\tilde r^{2}}\left(1+\frac{\mu}{\tilde r^{2}}+\frac{\mu a^{2}}{\tilde r^{4}}\right)h_{+-},
\\
Z_{33}&=2\left\{a\tilde r\left(1+\frac{\mu a^{2}}{\tilde r^{4}}\right)~\partial_{t}h_{3\tilde r}+\frac{2\mu a^{2}G(\tilde r)}{\tilde r^{2}}h_{\tilde r \tilde r}
-2\left(1+\frac{\mu a^{2}}{\tilde r^{4}}\right)h_{+-}\right\},\label{K=0:gaugeN:33}\\
Z_{\tilde r\tilde r}&=-\frac{2a}{\tilde r G(\tilde r)}\partial_{t}h_{\tilde r 3},\\
Z_{t \tilde r}&=-\frac{\mu a^{2}}{\tilde r^{3}}\left\{\partial_{t}h_{\tilde r\tilde r}
-\frac{1}{\tilde r^{3}G(\tilde r)}\partial_{\tilde r}\Big[\tilde r^{3}G(\tilde r) h_{t\tilde r}\Big]\right\}
-\frac{a}{\tilde r^{4}G(\tilde r)}\left(1-\frac{\mu}{\tilde r^{2}}\right)\partial_{\tilde r}\Big[\tilde r^{3}G(\tilde r) h_{\tilde r 3}\Big],\\
Z_{\tilde r 3}&=
a\tilde r\left(1+\frac{\mu a^{2}}{\tilde r^{4}}\right)\partial_{t}h_{\tilde r\tilde r}
-\frac{a\tilde r^{2}}{G(\tilde r)}\partial_{\tilde r}\Big[\tilde r^{-1} G(\tilde r) h_{t\tilde r}\Big]
-\frac{\mu a^{2}}{\tilde r^{4}}\frac{a}{\tilde r^{2}G(\tilde r)}\partial_{\tilde r}\Big[\tilde r^{3} G(\tilde r) h_{t\tilde r}\Big]
\notag\\&\qquad\qquad
-\frac{\mu a^{2}}{\tilde r^{6}G(\tilde r)}\partial_{\tilde r}\Big[\tilde r^{3}G(\tilde r)h_{\tilde r 3}\Big]
+2h_{\tilde r3},
\\
Z_{+-}&=2h_{+-},
\label{K=0:gaugeN:+-}
\end{align}
\end{subequations}
and the other components vanish.
Because the master variable for $Z_{\lambda\rho}$ is described by the gauge invariant variable that consists of $Z_{33}$ and $Z_{+-}$ as well as that for $h_{\lambda\rho}$, we focus on Eqs.~\eqref{K=0:gaugeN:33} and \eqref{K=0:gaugeN:+-}.
According to Ref.~\cite{Murata:2008yx}, the field equation for a linear metric perturbation $h_{\lambda\rho}$ includes $\delta G_{\tilde r3}=0$,
\begin{equation}
\frac{\mu a}{2\tilde r^{3}}\,\partial_{t}h_{\tilde r \tilde r}+\frac{1}{4 G(\tilde r)}\left(1+\frac{\mu a^{2}}{\tilde r^{4}}\right)\partial_{t}^{2}h_{\tilde r 3}-\frac{\mu a}{\tilde r^{5} G(\tilde r)}\partial_{t}h_{+-}=0.
\label{K=0:Gr3}
\end{equation}
Performing the integration of the both sides with respect to $t$ gives
\begin{equation}
\frac{\mu a}{2\tilde r^{3}}\,h_{\tilde r \tilde r}+\frac{1}{4 G(\tilde r)}\left(1+\frac{\mu a^{2}}{\tilde r^{4}}\right)\partial_{t}h_{\tilde r 3}-\frac{\mu a}{\tilde r^{5} G(\tilde r)}h_{+-}=0,
\label{K=0:Gr3:2}
\end{equation}
whence, applying Eq.~\eqref{K=0:Gr3:2} to Eq.~\eqref{K=0:gaugeN:33}, we arrive at
\begin{equation}
Z_{33}=-4\left(1-\frac{\mu a^{2}}{\tilde r^{4}}\right)h_{+-}.
\end{equation}
Then let us apply the same gauge condition as $h_{\lambda\rho}$ to $Z_{\lambda\rho}$.
The resulting linear metric perturbation $Z^{\rm g}_{\mu\nu}:=Z_{\mu\nu}+\delta Z_{\mu\nu}$ 
satisfies
\begin{equation}
Z^{\rm g}_{tt}=0,\quad
Z^{\rm g}_{t3}=0,\quad
Z^{\rm g}_{33}=0
\end{equation}
after the gauge transformation realized by the gauge field $\xi_{\mu}$ satisfying
\begin{equation}
\xi_{\tilde r}=-\Big(2\tilde r G(\tilde r)\Big)^{-1}\left(1-\frac{\mu a^{2}}{\tilde r^{4}}\right)^{-1}Z_{3 3}=4\Big(2\tilde r G(\tilde r)\Big)^{-1}h_{+-}.
\end{equation}
Simultaneously $Z^{\rm g}_{+-}$ results in
\begin{align}
Z^{\rm g}_{+-}&=Z_{+-}+2\tilde r G(\tilde r)\xi_{\tilde r}
=6h_{+-}.
\label{K=0:rel+-}
\end{align}
The master variable $\Phi^{Z}_{0}$ for $Z^{\rm g}_{\lambda\rho}$ is given in the same form as that for $h_{\lambda\rho}$, and therefore Eq.~\eqref{K=0:rel+-} immediately yields
\begin{equation}
\Phi^{Z}_{0}=6\Phi_{0}.
\label{K=0:PhiZ}
\end{equation}
In conclusion, the symmetry operator ${\cal S}$ for linear metric perturbations with $(J,K,M)=(0,0,0)$ 
maps the $(J,K,M)=(0,0,0)$ mode to itself, except the scaling by $6$.
%
It is worth noting that, since the map is trivial, the symmetry operator does not change the boundary behavior, 
and the regularity of a solution is also trivially preserved under the operation. 
For the $\omega=0$ case, we also find that the symmetry operator preserves 
the regularity of a solution as is shown in Appendix~\ref{app:stationary}.

\subsection{$(J,K,M)=(0,1,0)$ mode}

From Eq.~\eqref{hJKM}, the $(J,K,M)=(0,1,0)$ mode of a rank-2 symmetric tensor takes the form
\begin{equation}
h=h^{J=0}_{K=1\,M=0}+h^{J=0}_{K=-1\,M=0}=2h_{A+}(t,\tilde r)\bar\sigma^{A}\bar\sigma^{+}+2h_{A-}(t,\tilde r)\bar\sigma^{A}\bar\sigma^{-},
\label{K=1:h}
\end{equation}
where $A\in\{t,\tilde r,3\}$.
All the components are complex-valued, and they are set to $h_{A-}=h_{A+}^{*}$ so that $h_{\mu\nu}$ will remain real overall.
The gauge transformation $\delta h_{\mu\nu}=\nabla_{\mu}\xi_{\nu}+\nabla_{\nu}\xi_{\mu}$ for the $(J,K,M)=(0,1,0)$ mode with respect to a gauge field $\xi=\xi_{+}(t,\tilde r)\bar\sigma^{+}+\xi_{-}(t,\tilde r)\bar\sigma^{-}$ satisfying $\xi_{-}=\xi_{+}^{*}$ is calculated to be
\begin{subequations}
\begin{align}
\delta h_{t+}&=\partial_{t}\xi_{+}-\frac{2{\rm i}\mu a}{\tilde r^{4}}\xi_{+},\\
\delta h_{\tilde r +}&=\partial_{\tilde r}\xi_{+}-\frac{2}{\tilde r}\xi_{+},\\
\delta h_{+3}&=\frac{2{\rm i}\mu a^{2}}{\tilde r^{4}}\xi_{+}.
\end{align}
\end{subequations}
Then, we find that the following are gauge invariant:
\begin{subequations}
\begin{align}
f_{t}&:=\frac{1}{\tilde r^{2}}\left(h_{t+}+\frac{{\rm i}\tilde r^{4}}{2\mu a^{2}}\partial_{t}h_{+3}+a^{-1}h_{+3}\right),\label{ft}\\
f_{\tilde r}&:=\frac{1}{\tilde r^{2}}\left(h_{\tilde r +}+\frac{{\rm i}\tilde r^{2}}{2\mu a^{2}}\partial_{\tilde r}(\tilde r^{2} h_{+3})\right).
\label{fr}
\end{align}
\end{subequations}
Here, we should be careful that the terms including $h_{+3}$ of the gauge invariant variables $f_{t}$ and $f_{\tilde r}$ would diverge in the Schwarzschild case $a=0$.
In Appendix~\ref{app:C3}, we argue the Schwarzschild limit of the $(J,K,M)=(0,1,0)$ mode without introducing these gauge invariant variables.
During the following calculation, the combination
\begin{equation}
\partial_{t}f_{\tilde r}-\partial_{\tilde r}f_{t}=\partial_{t}\left(\tilde r^{-2}h_{\tilde r +}\right)-\partial_{\tilde r}\left(\tilde r^{-2}h_{t+}\right)-a^{-1}\partial_{\tilde r}\left(\tilde r^{-2}h_{+3}\right)
\end{equation}
derived from Eqs.~\eqref{ft} and \eqref{fr} will be 
frequently used. 
In detail, the master variable has been derived in Ref.~\cite{Murata:2008yx} as
\begin{equation}
\Phi_{1}:=\frac{\left(1+\frac{\mu a^{2}}{\tilde r^{4}}\right)^{1/4}}{\tilde r^{5/2}\left(1+\frac{\mu a^{2}}{\tilde r^{4}}+\frac{\mu^{2}a^{6}}{4\tilde r^{10}}\right)^{1/2}}~\pi_{\tilde r},
\label{Phi1}
\end{equation}
where $\pi_{\tilde r}$, which is (proportional to) the conjugate momentum to $f_{\tilde r}^{*}$, is defined as
\begin{equation}
\pi_{\tilde r}:=\tilde r^{5}\left(1+\frac{\mu a^{2}}{\tilde r^{4}}\right)\left(\partial_{t}f_{\tilde r}-\partial_{\tilde r}f_{t}\right)-2{\rm i} \mu a \tilde r f_{\tilde r}.
\label{K=1:pi}
\end{equation}

Substituting Eq.~\eqref{K=1:h} into Eqs.~\eqref{equal:ZAB}--\eqref{equal:Z+-}, we have $Z_{\lambda\rho}={\cal S}h_{\lambda\rho}$ for the $(J,K,M)=(0,1,0)$ mode as follows:
\begin{subequations}
\begin{align}
Z_{t+}&={\rm i}a \partial_{t}h_{t+}-\frac{\mu a^{2}}{\tilde r^{3}}\partial_{t}h_{\tilde r +}+\frac{\mu a^{2}}{\tilde r}\partial_{\tilde r}\left(\tilde r^{-2}h_{t+}\right)-\frac{a}{\tilde r^{3}}\left(1-\frac{\mu}{\tilde r^{2}}\right)\partial_{\tilde r}\left(\tilde r^{2}h_{+3}\right)
+2h_{t+}+\frac{2{\rm i}\mu a^{3}}{\tilde r^{5}}h_{\tilde r +}-\frac{4a}{\tilde r^{2}}\,h_{+3},
\label{K=1:Zt+}\\
Z_{\tilde r +}&={\rm i}a \partial_{t}h_{\tilde r +}-\frac{a}{\tilde r G(\tilde r)}\partial_{t}h_{+3}+\frac{2{\rm i}\mu a^{3}}{\tilde r^{5}G(\tilde r)}h_{t+}+2h_{\tilde r +}+\frac{2{\rm i}\mu a^{2}}{\tilde r^{5}G(\tilde r)}h_{+3},\label{K=1:Zr+}\\
Z_{+3}&=a\tilde r^{3}\left(1+\frac{\mu a^{2}}{\tilde r^{4}}\right)\Big\{\tilde r^{-2}\partial_{t}h_{\tilde r +}-\partial_{\tilde r}\left(\tilde r^{-2}h_{t+}\right)\Big\}
+{\rm i}a \partial_{t}h_{+3}
-\frac{\mu a^{2}}{\tilde r}\partial_{\tilde r}\left(\tilde r^{-2}h_{+3}\right)
-\frac{2{\rm i}\mu a^{2}}{\tilde r^{3}}h_{\tilde r +}+6h_{+3},\label{K=1:Z+3}
\end{align}
\end{subequations}
$Z_{A-}$ are given as the complex conjugate of $Z_{A+}$ as well as $h_{\mu\nu}$, and the other components vanish.
The gauge invariant variables $f_{t}^{Z}$ and $f_{\tilde r}^{Z}$ for $Z_{\lambda\rho}$ are given in the same form,
\begin{subequations}
\begin{align}
f_{t}^{Z}&:=\frac{1}{\tilde r^{2}}\left(Z_{t+}+\frac{{\rm i}\tilde r^{4}}{2\mu a^{2}}\partial_{t}Z_{+3}+a^{-1}Z_{+3}\right),\label{K=1:fNt}\\
f_{\tilde r}^{Z}&:=\frac{1}{\tilde r^{2}}\left(Z_{\tilde r +}+\frac{{\rm i}\tilde r^{2}}{2\mu a^{2}}\partial_{\tilde r}(\tilde r^{2} Z_{+3})\right),
\label{K=1:fNr}
\end{align}
\end{subequations}
as the gauge invariant variables $f_{t}$ and $f_{\tilde r}$, defined in Eqs.~\eqref{ft} and \eqref{fr} respectively.
The master variable $\Phi_{1}^{Z}$ for $Z_{\lambda\rho}$ is given in the same form as the master variable $\Phi_{1}^{Z}$ for $h_{\lambda\rho}$, given in Eq.~\eqref{Phi1}, and hence it includes
\begin{equation}
\pi_{\tilde r}^{Z}:=\tilde r^{5}\left(1+\frac{\mu a^{2}}{\tilde r^{4}}\right)\left(\partial_{t}f_{\tilde r}^{Z}-\partial_{\tilde r}f_{t}^{Z}\right)-2{\rm i} \mu a \tilde r f_{\tilde r}^{Z}
\label{K=1:piN}
\end{equation}
which is (proportional to) the conjugate momentum to $(f_{\tilde r}^{Z})^{*}$.

The expression of $\Phi_{1}^{Z}$ in terms of $\Phi_{1}$ is immediately obtained from the expression of $\pi_{\tilde r}^{Z}$ in terms of $\pi_{\tilde r}$.
Hence let us rewrite the right-hand side of Eq.~\eqref{K=1:piN}.
Substituting Eqs.~\eqref{K=1:Zt+}--\eqref{K=1:Z+3} into Eqs.~\eqref{K=1:fNt} and \eqref{K=1:fNr}, we arrive at the relations
\begin{subequations}
\begin{align}
f_{t}^{Z}&=\frac{{\rm i}\tilde r^{5}}{2\mu a}\left(1+\frac{\mu a^{2}}{\tilde r^{4}}\right)\partial_{t}(\partial_{t}f_{\tilde r}-\partial_{\tilde r}f_{t})
+\tilde r(\partial_{t}f_{\tilde r}-\partial_{\tilde r}f_{t})
+{\rm i}a\partial_{t}f_{t}
+\tilde r\partial_{t}f_{\tilde r}
+2f_{t}-\frac{2{\rm i}a}{\tilde r}\left(\frac{\mu}{\tilde r^{2}}-\frac{\mu a^{2}}{\tilde r^{4}}\right)f_{\tilde r},
\label{K=1:ft:new}
\\
f_{\tilde r}^{Z}&=\partial_{\tilde r}\left[\frac{{\rm i}\tilde r^{5}}{2\mu a}\left(1+\frac{\mu a^{2}}{\tilde r^{4}}\right)\left(\partial_{t}f_{\tilde r}-\partial_{\tilde r}f_{t}\right)\right]
+{\rm i}a\partial_{t}f_{\tilde r}
+\tilde r\partial_{\tilde r}f_{\tilde r}
+\frac{2{\rm i}\tilde r^{3}}{\mu a G(\tilde r)}\left(\frac{\mu a^{2}}{\tilde r^{4}}\right)^{2}f_{t}
+3f_{\tilde r}
\label{K=1:fr:new}
\end{align}
\end{subequations}
between $(f_{t},f_{\tilde r})$ and $(f_{t}^{Z},f_{\tilde r}^{Z})$,
and, from Eqs.~\eqref{K=1:ft:new} and \eqref{K=1:fr:new}, we also have
\begin{align}
\partial_{t}f_{\tilde r}^{Z}-\partial_{\tilde r}f_{t}^{Z}&=
{\rm i}a\,\partial_{t}(\partial_{t}f_{\tilde r}-\partial_{\tilde r}f_{t})
-\tilde r\partial_{\tilde r}(\partial_{t}f_{\tilde r}-\partial_{\tilde r}f_{t})+(\partial_{t}f_{\tilde r}-\partial_{\tilde r}f_{t})
\notag\\&\qquad 
+\frac{2{\rm i}\tilde r^{3}}{\mu a G(\tilde r)}\left(\frac{\mu a^{2}}{\tilde r^{4}}\right)^{2}\partial_{t}f_{t}
+\partial_{\tilde r}\left[\frac{2{\rm i}a}{\tilde r}\left(\frac{\mu}{\tilde r^{2}}-\frac{\mu a^{2}}{\tilde r^{4}}\right)f_{\tilde r}\right].
\label{K=1:rel}
\end{align}
We complete the calculation by substituting Eq.~\eqref{K=1:fr:new} and \eqref{K=1:rel} into Eq.~\eqref{K=1:piN}, applying the Euler-Lagrange equations for the system of $(f_{t},f_{\tilde r})$, and rewriting the all terms using Eq.~\eqref{K=1:pi}.
We present the details in Appendix~\ref{app:K=1}.
As a result, we have
\begin{equation}
\pi_{\tilde r}^{Z}=\left({\rm i}a\partial_{t}+6\right)\pi_{\tilde r},
\label{K=1:result}
\end{equation}
and therefore
\begin{equation}
\Phi_{1}^{Z}=\left({\rm i}a\partial_{t}+6\right)\Phi_{1}.
\label{K=1:result2}
\end{equation}
In conclusion, the symmetry operator ${\cal S}$ for linear metric perturbations with the $(J,K,M)=(0,1,0)$ is given as the combination 
of the identity and the phase shift, where we note that 
the imaginary part of the frequency induces the phase shift;
substituting $\Phi_{1}(t,\tilde r)={\rm e}^{-{\rm i}\omega t}\Phi_{1}(\tilde r)$ into Eq.~\eqref{K=1:result2} leads to
\begin{align}
\Phi_{1}^{Z}&=\left(a\omega+6\right)\Phi_{1}(t,\tilde r)
\notag\\
&=\Big\{\left(a\omega_{\rm Re}+6\right){\rm e}^{-{\rm i}\omega_{\rm Re} t}+a\omega_{\rm Im}{\rm e}^{-{\rm i}(\omega_{\rm Re} t-\frac{\pi}{2})}\Big\}{\rm e}^{\omega_{\rm Im}t}\Phi_{1}(\tilde r),
\end{align}
where $\omega=\omega_{\rm Re}+{\rm i}\omega_{\rm Im}$.

\subsection{$K=J+2$ modes}

From Eq.~\eqref{hJKM}, the $(J,K=J+2,M=0)$ mode of a rank-2 symmetric tensor takes the form
\begin{equation}
h=h^{J}_{K=J+2\,M}=h_{++}(t,\tilde r)\mathbb{D}^{J}_{K=J\,M=0}\,\bar\sigma^{+}\bar\sigma^{+}
+h_{--}(t,\tilde r)\mathbb{D}^{J}_{K=-J\,M=0}\,\bar\sigma^{-}\bar\sigma^{-},
\label{K=J+2:h}
\end{equation}
where $h_{--}$ is given as the complex conjugate of $h_{++}$, and consequently the master variable for $h$ is represented by $h_{++}$.

The non-zero components of $Z_{\lambda\rho}={\cal S}h_{\lambda\rho}$ are $Z_{++}$ and $Z_{--}=(Z_{++})^{*}$ as well as $h_{\lambda\rho}$.
From Eq.~\eqref{equal:Z++}, we have
\begin{equation}
Z_{++}=2\left({\rm i}a\partial_{t}+{\rm i}\bar\sigma_{3}+3\right)h_{++},
\label{K=J+2:Z}
\end{equation}
and therefore
\begin{equation}
\Phi_{K=J+2}^{Z}=2\left({\rm i}a\partial_{t}+{\rm i}\bar\sigma_{3}+3\right)\Phi_{K=J+2}.
\label{K=J+2:PhiZ}
\end{equation}
In conclusion, the symmetry operator ${\cal S}$ for linear metric perturbations with the $K=J+2$ mode is given as the combination of the identity and the phase shift,
where we note that 
the imaginary part of the frequency induces the phase shift;
substituting $\Phi_{K=J+2}={\rm e}^{-{\rm i}\omega t}\mathbb{D}^{J}_{K=J\,M=0}\Phi_{K=J+2}(\tilde r)$ into Eq.~\eqref{K=J+2:PhiZ}, 
and applying that ${\rm i}\bar\sigma_{3}=2W_{3}$, we obtain 
\begin{align}
\Phi_{K=J+2}^{Z}&=2\left(a\omega+2J+3\right)\Phi_{K=J+2}\notag\\
&=2\Big\{\left(a\omega_{\rm Re}+2J+3\right){\rm e}^{-{\rm i}\omega_{\rm Re} t}+a\omega_{\rm Im} {\rm e}^{-{\rm i}(\omega_{\rm Re} t-\frac{\pi}{2})}\Big\}{\rm e}^{\omega_{\rm Im} t}\mathbb{D}^{J}_{K=J\,M=0}\Phi_{K=J+2}(\tilde r),
\label{K=J+2:Phi}
\end{align}
where $\omega=\omega_{\rm Re}+{\rm i}\omega_{\rm Im}$.

\section{Summary and discussion}
\label{sec:V}

We investigated the symmetry operator constructed from the Killing-Yano 3-form for linear metric perturbations 
of the 5-dimensional Myers-Perry black hole spacetime with equal angular momenta.
We considered the Schwarzschild case and the finite angular momenta case.

In the Schwarzschild case, the symmetry operator eliminates the scalar perturbations, 
behaves as the operation $\hat{\star}\hat d$ to the vector perturbations, and behaves to the tensor perturbations similarly, 
where $\hat{\star}$ and $\hat d$ are the Hodge star operator and the exterior derivative on the unit 3-sphere, respectively. 
The vector harmonics have three classes $(\mathbb{V}^{(0)},\mathbb{V}^{(1)},\mathbb{V}^{(2)})$, 
and the tensor harmonics have six classes $(\mathbb{T}^{(0)},\mathbb{T}^{(1)},\mathbb{T}^{(2)},\mathbb{T}^{(3)},\mathbb{T}^{(4)},\mathbb{T}^{(5)})$
labeled by the same integer label $(\underline{k},\underline{\ell},\underline{m})$~\cite{doi:10.1063/1.523649,Lindblom:2017maa}.
The symmetry operator eliminates $\mathbb{V}^{(0)}$, $\mathbb{T}^{(0)}$ and $\mathbb{T}^{(3)}$, 
and maps each element of the pairs $(\mathbb{V}^{(1)},\mathbb{V}^{(2)})$, $(\mathbb{T}^{(1)},\mathbb{T}^{(2)})$ 
and $(\mathbb{T}^{(4)},\mathbb{T}^{(5)})$ 
to the other element of the pairs, except for multiplication by constant.
Thus the combinations $\mathbb{V}^{(1)}\pm\mathbb{V}^{(2)}$, $\mathbb{T}^{(1)}\pm\mathbb{T}^{(2)}$ 
and $\mathbb{T}^{(4)}\pm\mathbb{T}^{(5)}$ are eigenvectors and eigentensors of the symmetry operator, 
and the symmetry operator is regarded as the identity operator in terms of each eigenmode exhibited above.
Here, we note that we did not apply the field equation, that is, the result above applies to any rank-2 symmetric tensor 
in the 5-dimensional Schwarzschild spacetime.

In the finite angular momenta case, we carried out the mode decomposition of rank-2 symmetric tensors 
based on the group theoretical method of the $U(1)\times SU(2)$ isometry of the spacetime, 
following Refs.~\cite{Murata:2007gv,Murata:2008yx}.
The straightforward calculation showed that the symmetry operator 
maps 
a single mode to another tensor labeled by the same integer or half-integer label $(J,K,M)$.
This is originated from the commutativity of the symmetry operator ${\cal S}$ and 
the operators $({\cal W}_{3},{\cal L}_{i})$ from $U(1)\times SU(2)$.
Then, we considered linear metric perturbations by imposing the field equation, and analyzed the map by the symmetry operator in detail.
As a result, we observed that the symmetry operator can be regarded as the combination of the identity and the operators 
from the isometries of the spacetime; see the resulting master variables~\eqref{K=0:PhiZ}, \eqref{K=1:result2} and \eqref{K=J+2:PhiZ} 
for the $(J,K,M)=(0,0,0)$, $(J,K,M)=(0,1,0)$ and $K=J+2$ modes, respectively.

To summarize, through this paper, we gave an example of how the symmetry operator constructed from the Killing-Yano 3-form 
for linear metric perturbations works, where we showed that the symmetry operator commutes with those constructed from the isometries of 
the spacetime, and it is given as a linear combination of them with respect to each mode.
Although we did not investigate all modes, 
it would be reasonable to expect that the symmetry operator is given as the combination of 
the operators associated with the isometries for all modes in the finite angular momenta case.
As is mentioned in Sec.~\ref{sec:I}, our main interest was in the possibility to provide a ladder operator 
associated with the Killing-Yano 3-form. 
However, it turned out that the symmetry operator acts as the identity operator up to the isometries of the background spacetime 
for each mode investigated in this paper. 
One may be interested in how general is this statement: the hidden symmetry does not provide the associated ladder operator. 
Even if a general proof of commutativity would be hard to examine, 
we may perform additional checks in other spacetimes, such as the 7-dimensional Myers-Perry spacetime with enhanced symmetry, by applying similar calculations performed in this paper.
%
To the best of our knowledge, no study has focused on hidden symmetries as ladder operators for a certain mode decomposition of a field of arbitrary spin.
\footnote{It does never mean that there is no room for focusing on hidden symmetries as the ladder operators for modes.
One noteworthy example 
is the Laplace-Runge-Lenz vector in the Kepler problem that spans $SO(4)$ together with the angular momentum algebra.
Considering the motion as a geodesic in a curved spacetime through the Eisenhart lift, 
we can regard the Laplace-Runge-Lenz vector as the triplet of rank-2 Killing tensors on the Taub-NUT space (see, e.g. Ref.~\cite{Cariglia:2014ysa}). 
If we succeeded in constructing the symmetry operators for field equations from the symmetries of the space, 
they would be non-commutative.
}

We successfully carried out the evaluation of the map of rank-2 symmetric tensors 
for the reason that separation of variables and the mode decomposition was independently 
done due to the sufficient number of the isometries of the spacetime.
In much less symmetric spacetimes, we expect that the Killing-Yano 3-form could not contribute to the separability of the 
field equation for the following reasons.
We have explicitly shown that the symmetry operator just reduces to the identity or the isometry map 
in the given specific setting of this paper, and the result could be true for more general cases. 
We know from the experience in the Kerr spacetime that
the second order symmetry operator is most likely essential for the realization of the non-trivial separability.
Furthermore, even if we count the Killing-Yano 3-form in the symmetry operators which give 
the full separation of variables in addition to the isometries, 
the number of separation constants is not sufficient for realization of the separability in higher-dimensional spacetimes. 
Nevertheless, the operation of the symmetry operator associated with the Killing-Yano 3-form 
is not apparent at a glance, and further investigation is needed to understand a general 
property of the map with the symmetry operator associated with it.

\section{Acknowledgements}

This  work  was  supported  by  JSPS  KAKENHI  Grant  Number  JP19H01895, JP20H05850, JP20H05853 (C.Y.).

\appendix

\section{Derivation of Eqs.~(\ref{sch:rel1}), (\ref{sch:rel2}) and (\ref{sch:rel3})}
\label{app:Sch}

We derive the relations among the contributions in $Z_{\mu\nu}$ and $h_{\mu\nu}$ from scalar, vector and tensor perturbations summarized in Eqs.~\eqref{sch:rel1}, \eqref{sch:rel2} and \eqref{sch:rel3}.
We will apply the formula
\begin{equation}
\hat D_{m}\hat\varepsilon_{ikl}=0
\end{equation}
for the volume form $\hat\varepsilon_{ijk}$ and the covariant derivative $\hat D_{i}$ 
associated with $\gamma_{ij}
$, and we also use that the Riemann curvature tensor $\hat R_{ijkl}$ 
associated with $\hat D_{i}$ for $S^{3}$, a space of constant curvature, is given by
\begin{equation}
\hat R_{ijkl}=\gamma_{ik}\gamma_{jl}-\gamma_{il}\gamma_{jk}.
\label{const}
\end{equation}

\subsection{$(a,i)$ components}

The expression of $Z_{ai}$ in terms of $h_{ai}$ has been given in Eq.~\eqref{ai}, and the decomposition of $Z_{ai}$ and $h_{ai}$ into the contributions from scalar and vector perturbations has been given in Eqs.~\eqref{hai:decomp} and \eqref{Zai:decomp}.
Calculating the divergence of $Z_{ai}$ 
associated with $\gamma_{ij}$ yields
\begin{align}
\hat D^{i}Z_{ai}&=\hat\varepsilon_{i}{}^{kl}\left(\hat D^{i}\hat D_{k}h_{al}\right)
=\frac{1}{2}\hat\varepsilon_{i}{}^{kl}\left([\hat D^{i},\hat D_{k}]h_{al}\right)
=\frac{1}{2}\hat\varepsilon_{i}{}^{kl}\left(\hat R_{l}{}^{mi}{}_{k}h_{am}\right)=0.
\end{align}
Note that we applied the expression~\eqref{const} of $\hat R_{ijkl}$ in the last equality.
This implies that $Z_{a}=0$.
Then, evaluating $Z_{ai}[h_{a}]$ gives
\begin{align}
Z_{ai}[h_{a}]&=\hat\varepsilon_{i}{}^{kl}\hat D_{k}\hat D_{l}h_{a}=0.
\end{align}
Therefore the only non-vanishing part of $Z_{ai}$ is $Z_{ai}=Z_{T}\,_{ai}[h_{T}\,_{ai}]$.

\subsection{$(i,j)$ components}

The expression of $Z_{ij}$ in terms of $h_{ij}$ has been given in Eq.~\eqref{ij}, and the decomposition of $Z_{ij}$ and $h_{ij}$ to 
the scalar, vector and tensor contributions has been given in Eqs.~\eqref{hij:decomp} and \eqref{Zij:decomp}, respectively.
The trace of $Z_{ij}$ is calculated to be
\begin{equation}
\gamma^{ij}Z_{ij}=0.
\end{equation}
Therefore $Z_{L}=0$.
Substituting $h_{ij}=h_{L}\gamma_{ij}$ into $Z_{ij}$ leads to
\begin{align}
Z_{ij}[h_{L}]&=\hat\varepsilon_{i}{}^{kl}\hat D_{k}\left(h_{L}\gamma_{jl}\right)+\hat\varepsilon_{j}{}^{kl}\hat D_{k}\left(h_{L}\gamma_{il}\right)\notag\\
&=-\hat\varepsilon_{ij}{}^{l}\hat D_{l}h_{L}+\hat\varepsilon_{ij}{}^{l}\hat D_{l}h_{L}=0.
\end{align}
Substituting $h_{ij}=\hat L_{ij}h_{T}$ into $Z_{ij}$ leads to
\begin{align}
Z_{ij}[h_{T}]&=2\varepsilon_{(i}{}^{kl}\hat D_{|k|}\left(\hat L_{j)l}h_{T}\right)\notag\\
&=2\varepsilon_{(i}{}^{kl}\hat D_{|k|}\left[\left(\hat D_{j)}\hat D_{l}-\frac{1}{3}\gamma_{j)l}\hat \triangle\right)h_{T}\right]\notag\\
&=\varepsilon_{(i}{}^{kl}[\hat D_{k},\hat D_{l}]\hat D_{j)}h_{T}\notag\\
&=\varepsilon_{(i}{}^{kl}\hat R_{j)}{}^{m}{}_{kl}\hat D_{m}h_{T}
=0.
\end{align}
So far, the possible contributions of $h_{ij}$ to $Z_{ij}$ are 
$Z_{ij}=Z_{ij}[h_{T}\,{}_{ij}]+Z_{ij}[h_{T}\,{}_{i}]$.
The divergence of $Z_{ij}$ with respect to the first index, 
associated with $\gamma_{ij}$, is calculated to be
\begin{align}
\hat D^{i}Z_{ij}&=2\hat D^{i}\left(\hat\varepsilon_{(i}{}^{kl}\hat D_{|k|}h_{j)l}\right)\notag\\
&=\frac{1}{2}\hat\varepsilon_{i}{}^{kl}[\hat D^{i},\hat D_{k}]h_{jl}+\hat\varepsilon_{j}{}^{kl}[\hat D^{i},\hat D_{k}]h_{il}+\hat\varepsilon_{j}{}^{kl}\hat D_{k}\hat D^{i}h_{il}\notag\\
&=\frac{1}{2}\hat\varepsilon_{i}{}^{kl}\left(\hat R_{j}{}^{mi}{}_{k}h_{ml}+\hat R_{l}{}^{mi}{}_{k}h_{jm}\right)
+\hat\varepsilon_{j}{}^{kl}\left(\hat R_{i}{}^{mi}{}_{k}h_{ml}+\hat R_{l}{}^{mi}{}_{k}h_{im}\right)+\hat\varepsilon_{j}{}^{kl}\hat D_{k}\hat D^{i}h_{il}\notag\\
&=\hat\varepsilon_{j}{}^{kl}\hat D_{k}\hat D^{i}h_{il}.
\end{align}
This immediately yields that $\hat D^{i}Z_{ij}[h_{T}\,{}_{ij}]=0$, and therefore $\hat D^{i}Z_{ij}$ admits only the contribution of $h_{T}\,{}_{i}$:
\begin{align}
\hat D^{i}Z_{ij}=
\hat D^{i}Z_{ij}[h_{T}\,{}_{i}]
&=\hat\varepsilon_{j}{}^{kl}\hat D_{k}\hat D^{i}\left(2\hat D_{(i}h_{T}\,{}_{l)}\right)\notag\\
&=\hat\varepsilon_{j}{}^{kl}\hat D_{k}\hat D^{i}\left(\hat D_{i}h_{T}\,{}_{l}+\hat D_{l}h_{T}\,{}_{i}\right)\notag\\
&=\hat\varepsilon_{j}{}^{kl}\hat D_{k}\Big(\hat D^{i}\hat D_{i}h_{T}\,{}_{l}+\frac{\hat R}{3}h_{T}\,{}_{l}\Big).
\label{DZ}
\end{align}
From Eq.~\eqref{DZ}, we also find that $\hat D^{j}\hat D^{i}Z_{ij}$ vanishes:
\begin{align}
\hat D^{j}\hat D^{i}Z_{ij}
&=\hat\varepsilon_{j}{}^{kl}\hat D^{j}\hat D_{k}\Big(\hat D^{i}\hat D_{i}h_{T}\,{}_{l}+\frac{\hat R}{3}h_{T}\,{}_{l}\Big)\notag\\
&=\frac{1}{2}\hat\varepsilon_{j}{}^{kl}[\hat D^{j},\hat D_{k}]\Big(\hat D^{i}\hat D_{i}h_{T}\,{}_{l}+\frac{\hat R}{3}h_{T}\,{}_{l}\Big)\notag\\
&=\frac{1}{2}\hat\varepsilon_{j}{}^{kl}\hat R_{l}{}^{mj}{}_{k}\Big(\hat D^{i}\hat D_{i}h_{T}\,{}_{m}+\frac{\hat R}{3}h_{T}\,{}_{m}\Big)
=0.
\label{DD1}
\end{align}
On the other hand, in terms the decomposition~\eqref{Zij:decomp} of $Z_{ij}$ together with the properties~\eqref{property:Z}, we have $\hat D^{j}\hat D^{i}Z_{T}\,{}_{ij}=0$ and
\begin{align}
\hat D^{j}\hat D^{i}\left(2\hat D_{(i}Z_{T}\,{}_{j)}\right)
&=\hat D^{j}\hat D^{i}\left(\hat D_{i}Z_{T}\,{}_{j}+\hat D_{j}Z_{T}\,{}_{i}\right)\notag\\
&=\left\{\left([\hat D^{j},\hat D^{i}]\hat D_{i}+\hat D^{i}[\hat D^{j},\hat D_{i}]\right)Z_{T}\,{}_{j}+\left(D^{j}[\hat D^{i},\hat D_{j}]\right)Z_{T}\,{}_{i}\right\}
=0.
\end{align}
Therefore $\hat D^{j}\hat D^{i}Z_{ij}$ includes only the contribution of $Z_{T}$:
\begin{equation}
\hat D^{j}\hat D^{i}Z_{ij}=\hat D^{j}\hat D^{i}\left(\hat L_{ij}Z_{T}\right).
\label{DD2}
\end{equation}
Comparing Eqs.~\eqref{DD1} and \eqref{DD2}, we have $Z_{T}=0$.
Then, let us return to $\hat D^{i}Z_{ij}$.
Eq.~\eqref{DD1} is further rewritten as
\begin{align}
\hat D^{i}Z_{ij}
&=\hat\varepsilon_{j}{}^{kl}\hat D_{k}\Big(\hat D^{i}\hat D_{i}h_{T}\,{}_{l}+\frac{\hat R}{3}h_{T}\,{}_{l}\Big)\notag\\
&=\hat\varepsilon_{j}{}^{kl}\Big\{\Big([\hat D_{k},\hat D^{i}]\hat D_{i}+\hat D^{i}[\hat D_{k},\hat D_{i}]+\hat D^{i}\hat D_{i}\hat D_{k}\Big)h_{T}\,{}_{l}+\frac{\hat R}{3}\hat D_{k}h_{T}\,{}_{l}\Big\}\notag\\
&=\hat\varepsilon_{j}{}^{kl}\Big\{\Big([\hat D_{k},\hat D^{i}]\hat D_{i}+\hat D^{i}[\hat D_{k},\hat D_{i}]\Big)h_{T}\,{}_{l}+\Big(\hat D^{i}\hat D_{i}+\frac{\hat R}{3}\Big)\hat D_{k}h_{T}\,{}_{l}\Big\},
\end{align}
where we find that the first term vanishes:
\begin{align}
\hat\varepsilon_{j}{}^{kl}\Big([\hat D_{k},\hat D^{i}]\hat D_{i}+\hat D^{i}[\hat D_{k},\hat D_{i}]\Big)h_{T}\,{}_{l}
&=\hat\varepsilon_{j}{}^{kl}\Big\{-\hat R^{m}{}_{k}\hat D_{m}h_{T}\,{}_{l}+\hat R_{l}{}^{m}{}_{k}{}^{i}\hat D_{i}h_{T}\,{}_{m}+\hat D^{i}\Big(\hat R_{l}{}^{m}{}_{ki}h_{T}\,{}_{m}\Big)\Big\}\notag\\
&=\hat\varepsilon_{j}{}^{kl}\Big(-1+\frac{1}{2}+\frac{1}{2}\Big)\frac{\hat R}{3}\hat D_{k}h_{T}\,{}_{l}
=0.
\end{align}
We have
\begin{equation}
\hat D^{i}Z_{ij}=\hat\varepsilon_{j}{}^{kl}\Big(\hat D^{i}\hat D_{i}+\frac{\hat R}{3}\Big)\hat D_{k}h_{T}\,{}_{l}.
\label{ijDh}
\end{equation}
On the other hand, from the divergenceless property $\hat D^{i}Z_{T}\,{}_{ij}=0$ of $Z_{T}\,{}_{ij}$, given in Eq.~\eqref{property:Z}, we find that $\hat D^{i}Z_{ij}$ includes only the contribution of $Z_{T}\,{}_{i}$:
\begin{align}
\hat D^{i}Z_{ij}&=\hat D^{i}\left(2\hat D_{(i}Z_{T}\,{}_{j)}\right)\notag\\
&=\hat D^{i}\hat D_{i}Z_{T}\,{}_{j}+[\hat D^{i},\hat D_{j}]Z_{T}\,{}_{i}\notag\\
&=\hat D^{i}\hat D_{i}Z_{T}\,{}_{j}+\frac{\hat R}{3}Z_{T}\,{}_{j}.
\label{ijDZ}
\end{align}
Comparing Eqs.~\eqref{ijDh} and \eqref{ijDZ}, we can set
$Z_{T}\,{}_{j}=\hat\varepsilon_{j}{}^{kl}\hat D_{k}h_{T}\,{}_{l}$.
Finally, we have
\begin{align}
Z_{T}\,{}_{ij}&=Z_{ij}-2\hat D_{(i}Z_{T}\,{}_{j)}\notag\\
&=2\varepsilon_{(i}{}^{kl}\hat D_{|k|}\left(h_{T}\,{}_{j)l}+\hat D_{j)}h_{T}\,{}_{l}+\hat D_{l}h_{T}\,{}_{j)}\right)-2\hat\varepsilon_{(i}{}^{kl}\hat D_{j)}\hat D_{k}h_{T}\,{}_{l}\notag\\
&=2\varepsilon_{(i}{}^{kl}\hat D_{|k|}h_{T}\,{}_{j)l}+\varepsilon_{(i}{}^{kl}[\hat D_{k},\hat D_{l}]h_{T}\,{}_{j)}-2\hat\varepsilon_{(i}{}^{kl}[\hat D_{j)},\hat D_{k}]h_{T}\,{}_{l}\notag\\
&=2\varepsilon_{(i}{}^{kl}\hat D_{|k|}h_{T}\,{}_{j)l}.
\end{align}

\section{Stationary $(J,K,M)=(0,0,0)$ mode}
\label{app:stationary}

We present the action of the 
symmetry operater on the 
stationary $(J,K,M)=(0,0,0)$ mode of the linear metric perturbation.  

Since the gauge fixing~\eqref{K=0:fix} does not work 
without time dependence of the components of the metric perturbation, 
we adopt the other gauge condition given by
\begin{equation}
h^{\rm g}_{t\tilde r}=0,\quad h^{\rm g}_{\tilde r 3}=0,\quad h^{\rm g}_{+-}=0. 
\end{equation}
For the gauge transformation associated with the gauge field 
$\xi=\xi_{\tilde r}(\tilde r)\bar\sigma^{\tilde r}+t[\xi_{t}(\tilde r)\bar\sigma^{t}+\xi_{3}(\tilde r)\bar\sigma^{3}]$, 
the component $\xi_{\tilde r}$ is uniquely determined to
\begin{equation}
\xi_{\tilde r}=-\frac{h_{+-}}{2\tilde rG(\tilde r)},
\end{equation} 
and the other components $\xi_{t}$ and $\xi_{3}$ are solutions of the simultaneous differential equations
\begin{subequations}
\begin{align}
\partial_{\tilde r}\xi_{t}-\frac{2\mu}{\tilde r^{3}G(\tilde r)}\xi_{t}-\frac{2\mu a}{\tilde r^{5} G(\tilde r)}\xi_{3}&=-h_{t\tilde r},\\
\partial_{\tilde r}\xi_{3}+\frac{4\mu a}{\tilde r^{3}G(\tilde r)}\xi_{t}-\frac{2}{\tilde r G(\tilde r)}\left(1-\frac{\mu}{\tilde r^{2}}-\frac{\mu a^{2}}{\tilde r^{4}}\right)\xi_{3}&=-h_{\tilde r3}. 
\end{align}
\end{subequations}
These differential equations allow the residual degrees of gauge freedom associated with 
the superposition of homogeneous solutions. 
Then the residual gauge transformations which preserve the stationarity of the metric components are 
restricted to the following two degrees of freedom: 
\begin{equation}
\delta h_{tt}\propto-2\left(1-\frac{\mu}{\tilde r^{2}}\right),\quad
\delta h_{t3}\propto-\frac{\mu a}{\tilde r^{2}},
\label{eq:g1}
\end{equation}
and 
\begin{equation}
\delta h_{tt}\propto-\frac{2\mu a}{\tilde r^{2}},\quad
\delta h_{t3}\propto\tilde r^{2}\left(1+\frac{\mu a^{2}}{\tilde r^{4}}\right). 
\label{eq:g2}
\end{equation}
These variations of the metric components correspond to the coordinate transformations 
$t\mapsto (1+\epsilon) t$ and $\tilde \phi\mapsto\tilde\phi+\epsilon t$. 
Hereafter, we omit the superscript ``~${}^{\rm g}$~'' of the resulting perturbation for convenience.

The field equation $\delta G_{\mu\nu}=0$ for the perturbation includes five non-zero differential equations
\begin{equation}
\delta G_{AB}=0,\quad \delta G_{\tilde r\tilde r}=0,\quad \delta G_{+-}=0,
\label{eq:stationary}
\end{equation}
where one equation is redundant. 
Then we obtain four equations 
for the four components $h_{AB}(\tilde r)$ and $h_{\tilde r\tilde r}(\tilde r)$, where $A, B\in\{t,3\}$.
By performing the deformation of the simultaneous equations~\eqref{eq:stationary}, we have three second order differential equations for $h_{AB}(\tilde r)$ whose general solution includes six integral constants, and we also have the equation determining $h_{\tilde r\tilde r}(\tilde r)$: 
\begin{align}
h_{\tilde r\tilde r}
&=\frac{1}{2\left(3-\frac{\mu a^{2}}{\tilde r^{4}}\right)[\tilde rG(\tilde r)]^{2}}
\Big\{
-\tilde r G(\tilde r)\left[\tilde r^{2}\left(3+\frac{\mu a^{2}}{\tilde r^{4}}\right)\partial_{\tilde r}h_{tt}+\frac{2\mu a}{\tilde r^{2}}\partial_{\tilde r}h_{t3}-\left(2-\frac{\mu}{\tilde r^{2}}\right)\partial_{\tilde r} h_{33}\right]\Big.
\notag\\
&\qquad\qquad
+2\mu\left(3-\frac{\mu a^{2}}{\tilde r^{4}}\right)h_{tt}
+\frac{4\mu a}{\tilde r^{2}}\left(4-\frac{\mu}{\tilde r^{2}}\right)h_{t3}
-2\left[\left(1-\frac{\mu}{\tilde r^{2}}\right)^{2}-\frac{3\mu a^{2}}{\tilde r^{4}}\right]h_{33}
\Big\}.
\end{align}
A straightforward calculation shows that the general solution consists of the following six contributions, 
and two among them are given by the residual gauge transformations \eqref{eq:g1} and \eqref{eq:g2}. 
The other four contributions are the following physical solutions of the perturbation equations: 
the perturbation $h^{\rm md}$ to the different mass parameter $\mu\mapsto\mu+\epsilon$ whose components are
\begin{equation}
h^{\rm md}_{tt}=\frac{1}{\tilde r^{2}},\quad
h^{\rm md}_{t3}=-\frac{a}{\tilde r^{2}},\quad
h^{\rm md}_{33}=\frac{a^{2}}{\tilde r^{2}},\quad
h^{\rm md}_{\tilde r\tilde r}=\frac{\tilde r^{4}(\tilde r^{2}-a^{2})}{\tilde r^{8}G^{2}},
\end{equation}
the perturbation $h^{\rm sd}$ to the different equal spin parameters $\mu a\mapsto \mu a+\epsilon$ whose components are
\begin{equation}
h^{\rm sd}_{t3}=-\frac{1}{\tilde r^{2}},\quad
h^{\rm sd}_{33}=\frac{2a}{\tilde r^{2}},\quad
h^{\rm sd}_{\tilde r\tilde r}=-\frac{2a}{\tilde r^{4}G(\tilde r)^{2}},
\end{equation}
the squashing $h^{\rm sq}$ along the Dobiasch-Maison family~\cite{Dobiasch:1981vh} whose components are
\begin{gather}
h^{\rm sq}_{tt}=\mu\left(1-\frac{\mu}{2\tilde r^{2}}\right),\quad
h^{\rm sq}_{t3}=-\mu a\left(1-\frac{\mu}{2\tilde r^{2}}\right),\quad
h^{\rm sq}_{33}=-\tilde r^{4}\left(1-\frac{\mu a^{2}}{\tilde r^{4}}\right)\left(1-\frac{\mu}{2\tilde r^{2}}\right),
\notag\\
h^{\rm sq}_{\tilde r\tilde r}=\frac{\tilde r^{2}}{G(\tilde r)^{2}}\left\{-2G(\tilde r)+\left(1-\frac{\mu a^{2}}{\tilde r^{4}}\right)\left(1-\frac{\mu}{2\tilde r^{2}}\right)\right\},
\end{gather}
and the non-physical one $h^{\rm si}$ that is singular on the horizon whose components are
\begin{gather}
h^{\rm si}_{tt}=-3D+\frac{D^{2}}{\mu}\left(1-\frac{\mu}{2\tilde r^{2}}\right)\ln N(\tilde r),\quad
h^{\rm si}_{t3}=-\frac{3a D}{2}+\left(-\frac{3\mu a}{4}+\frac{\mu a(\mu+2a^{2})}{8\tilde r^{2}}+\frac{3a\tilde r^{2}}{4}\right)\ln N(\tilde r),\notag\\
h^{\rm si}_{33}=-\frac{3}{2}D \tilde r^{2}+\left(\frac{3\mu a^{2}}{4}-\frac{\mu a^{2}(\mu+8a^{2})}{8\tilde r^{2}}+\frac{3\mu\tilde r^{2}}{8}-\frac{3\tilde r^{4}}{4}\right)\ln N(\tilde r),
\\
h^{\rm si}_{\tilde r\tilde r}=\frac{\tilde r^{2}}{8G(\tilde r)^{2}}\left\{\frac{2D}{\tilde r^{4}}\left(\frac{4D^{2}}{\mu}+3\tilde r^{2}+3\mu\right)
+\left[6-\frac{9\mu}{\tilde r^{2}}+\frac{2\mu}{\tilde r^{4}} (\mu+5a^{2})-\frac{\mu a^{2}}{\tilde r^{6}}(\mu+8 a^{2})\right]\ln N(\tilde r)\right\},\notag
\end{gather}
where 
\begin{equation}
N(\tilde r)=\frac{M-D+\tilde r^{2}}{M+D+\tilde r^{2}},\quad
D=\frac{1}{2}\sqrt{\mu(\mu-4a^{2})},\quad
M=\frac{\mu}{2}.
\end{equation}
Note that the horizon radii $r_{\pm}$ are given by $r_{\pm}^{2}=M\pm D$, and one can confirm that the singular solution $h^{\rm si}$ diverges on the horizon.

Finally, we show the action of the symmetry operator on 
each contribution to the stationary perturbation. 
As a result, we have
\begin{subequations}
\begin{align}
{\cal S}[h^{\rm g1}]&=0,\\
{\cal S}[h^{\rm g2}]&=4ah^{\rm g1}+2h^{\rm g2},\\
{\cal S}[h^{\rm md}]&=-ah^{\rm sd},\\
{\cal S}[h^{\rm sd}]&=2h^{\rm sd},\\
{\cal S}[h^{\rm sq}]&=6\left\{h^{\rm sq}+\frac{\mu}{3}h^{\rm g1}+\frac{2a}{3}h^{\rm g2}-\frac{4D^{2}}{3} h^{\rm md}+\frac{2aD^{2}}{3} h^{\rm sd}\right\},\\
{\cal S}[h^{\rm si}]&=6\left\{h^{\rm si}-\frac{3D}{2}h^{\rm g1}+\frac{D}{2}(7\mu-4a^{2})h^{\rm md}-\frac{aD^{3}}{\mu}h^{\rm sd}\right\},
\end{align}
\end{subequations}
where $h^{\rm g1}$ and $h^{\rm g2}$ are the residual gauge transformation, given in Eqs.~\eqref{eq:g1} and \eqref{eq:g2}.
We can obtain six eigentensors for the symmetry operator by considering the proper linear combination of them.

\section{Derivation of Eq.~(\ref{K=1:result})}
\label{app:K=1}

We present the canonical equations for the system of $(f_{t},f_{\tilde r})$, which have been derived in Ref.~\cite{Murata:2008yx}, in Appendix~\ref{app:K=1:can}, and then we apply the canonical equations to $\pi_{\tilde r}^{Z}$, given in Eq.~\eqref{K=1:piN}, to obtain the expression~\eqref{K=1:result} of $\pi_{\tilde r}^{Z}$ in terms of $\pi_{\tilde r}$ in Appendix~\ref{app:K=1:der}.

\subsection{Hamiltonian system}
\label{app:K=1:can}

The canonical equations for the system of $(f_{t},f_{\tilde r})$ have been derived in Appendix~B of Ref.~\cite{Murata:2008yx}.
Following Ref.~\cite{Murata:2008yx}, we introduce the notations
\footnote
{
${\rm e}^{\mathscr{D}}$ and ${\rm e}^{\mathscr{E}}$ can be negative, depending on the value of $a$, and hence the current notations are inappropriate in this sense.
However, this point is not important, and we prioritize the consistency with Ref.~\cite{Murata:2008yx}.
}
\begin{gather}
{\rm e}^{\mathscr{A}}:=\tilde r^{5}\left(1+\frac{\mu a^{2}}{\tilde r^{4}}\right),\quad
{\rm e}^{\mathscr{B}}:=\frac{4\tilde r^{3}}{G(\tilde r)}\left(\frac{\mu a^{2}}{\tilde r^{4}}\right)^{2},\quad
{\rm e}^{\mathscr{C}}:=\frac{4\mu a^{2}}{\tilde r}\left(\frac{\mu}{\tilde r^{2}}-\frac{\mu a^{2}}{\tilde r^{4}}\right),
\notag\\
{\rm e}^{\mathscr{D}}:=-2\mu a \tilde r,\quad
{\rm e}^{\mathscr{E}}:=-8\mu a,
\end{gather}
and the action for the $(J,K,M)=(0,1,0)$ mode is given as
\begin{align}
S&\propto\int dtd\tilde r\,\left\{{\rm e}^{\mathscr{A}}|\partial_{t}f_{\tilde r}-\partial_{\tilde r}f_{t}|^{2}+{\rm e}^{\mathscr{B}}|f_{t}|^{2}
+{\rm e}^{\mathscr{C}}|f_{\tilde r}|^{2}
+2\,{\rm Im}\Big[{\rm e}^{\mathscr{D}}(\partial_{t}f_{\tilde r}-\partial_{\tilde r}f_{t})f_{\tilde r}^{*}-{\rm e}^{\mathscr{E}} f_{t}f_{\tilde r}^{*}\Big]
\right\}.
\end{align}
We can define the conjugate momentum $\pi_{\tilde r}$ to $f_{\tilde r}^{*}$ as Eq.~\eqref{K=1:pi}, which can be rewritten as
\begin{equation}
\pi_{\tilde r}={\rm e}^{\mathscr{A}}(\partial_{t}f_{\tilde r}-\partial_{\tilde r}f_{t})+{\rm i}\,{\rm e}^{\mathscr{D}}f_{\tilde r},
\label{K=1:pi:e}
\end{equation}
and we can also derive the constraint equation for $f_{t}$ and the canonical equations for $(f_{\tilde r},\pi_{\tilde r})$ as follows:
\begin{subequations}
\begin{gather}
{\rm e}^{\mathscr{B}}f_{t}=-\partial_{\tilde r}\pi_{\tilde r}+{\rm i}\,{\rm e}^{\mathscr{E}}f_{\tilde r},
\label{K=1:eom1}
\\
\partial_{t}f_{\tilde r}+{\rm i}\,{\rm e}^{\mathscr{D-A}}f_{\tilde r}={\rm e}^{-\mathscr{A}}\pi_{\tilde r}-\partial_{\tilde r}({\rm e}^{-\mathscr{B}}\partial_{\tilde r}\pi_{\tilde r})+{\rm i}\partial_{\tilde r}({\rm e}^{\mathscr{E-B}}f_{\tilde r}),
\label{K=1:eom2}
\\
\partial_{t}\pi_{\tilde r}=-({\rm e}^{2\mathscr{E-B}}-{\rm e}^{\mathscr{C}}+{\rm e}^{2\mathscr{D-A}})f_{\tilde r}-{\rm i}\,{\rm e}^{\mathscr{E-B}}\partial_{\tilde r}\pi_{\tilde r}-{\rm i}\,{\rm e}^{\mathscr{D-A}}\pi_{\tilde r}.
\label{K=1:eom3}
\end{gather}
\end{subequations}

\subsection{Derivation of Eq.~\eqref{K=1:result}}
\label{app:K=1:der}

We can rewrite the conjugate momentum $\pi_{\tilde r}^{Z}$ to $(f_{\tilde r}^{Z})^{*}$, given in Eq.~\eqref{K=1:piN}, as
\begin{equation}
\pi_{\tilde r}^{Z}={\rm e}^{\mathscr{A}}\left(\partial_{t}f_{\tilde r}^{Z}-\partial_{\tilde r}f_{t}^{Z}\right)+{\rm i}\,{\rm e}^{\mathscr{D}}f_{\tilde r}^{Z}
\label{K=1:piN:e}
\end{equation}
using the above notations as well as $\pi_{\tilde r}$. 
Eqs.~\eqref{K=1:ft:new}, \eqref{K=1:fr:new} and \eqref{K=1:rel} are likewise rewritten as
\begin{subequations}
\begin{align}
f_{t}^{Z}&=-4{\rm i}\,{\rm e}^{\mathscr{A-E}}\partial_{t}(\partial_{t}f_{\tilde r}-\partial_{\tilde r}f_{t})
+\tilde r(\partial_{t}f_{\tilde r}-\partial_{\tilde r}f_{t})
+{\rm i}a\partial_{t}f_{t}
+\tilde r\partial_{t}f_{\tilde r}
+2f_{t}+4{\rm i}\,{\rm e}^{\mathscr{C-E}}f_{\tilde r},\\
f_{\tilde r}^{Z}&=-4{\rm i}\,{\rm e}^{-\mathscr{E}}\partial_{\tilde r}\left[e^{\mathscr{A}}\left(\partial_{t}f_{\tilde r}-\partial_{\tilde r}f_{t}\right)\right]
+{\rm i}a\partial_{t}f_{\tilde r}
+\tilde r\partial_{\tilde r}f_{\tilde r}
-4{\rm i}\,{\rm e}^{\mathscr{B-E}}f_{t}
+3f_{\tilde r},
\label{K=1:frN:e}
\end{align}
\end{subequations}
and
\begin{equation}
\partial_{t}f_{\tilde r}^{Z}-\partial_{\tilde r}f_{t}^{Z}=
{\rm i}a\,\partial_{t}(\partial_{t}f_{\tilde r}-\partial_{\tilde r}f_{t})
-\tilde r\partial_{\tilde r}(\partial_{t}f_{\tilde r}-\partial_{\tilde r}f_{t})+(\partial_{t}f_{\tilde r}-\partial_{\tilde r}f_{t})
-4{\rm i}\,{\rm e}^{\mathscr{B-E}}\partial_{t}f_{t}
-\partial_{\tilde r}\left[4{\rm i}\,{\rm e}^{\mathscr{C-E}}f_{\tilde r}\right].
\label{K=1:rel:e}
\end{equation}
Let us rewrite Eq.~\eqref{K=1:piN:e} using the canonical equations~\eqref{K=1:eom1}--\eqref{K=1:eom3}.
First, substituting Eqs.~\eqref{K=1:frN:e} and \eqref{K=1:rel:e} into Eq.~\eqref{K=1:piN:e} gives
\begin{align}
\pi_{\tilde r}^{Z}&={\rm e}^{\mathscr{A}}\Big\{{\rm i}a\,\partial_{t}(\partial_{t}f_{\tilde r}-\partial_{\tilde r}f_{t})
-\tilde r\partial_{\tilde r}(\partial_{t}f_{\tilde r}-\partial_{\tilde r}f_{t})+(\partial_{t}f_{\tilde r}-\partial_{\tilde r}f_{t})
-4{\rm i}\,{\rm e}^{\mathscr{B-E}}\partial_{t}f_{t}
-\partial_{\tilde r}\left[4{\rm i}\,{\rm e}^{\mathscr{C-E}}f_{\tilde r}\right]\Big\}\notag\\
&\quad-{\rm i}\,{\rm e}^{\mathscr{D}}\Big\{4{\rm i}\,{\rm e}^{-\mathscr{E}}\partial_{\tilde r}\left[{\rm e}^{\mathscr{A}}\left(\partial_{t}f_{\tilde r}-\partial_{\tilde r}f_{t}\right)\right]
-{\rm i}a\partial_{t}f_{\tilde r}
-\tilde r\partial_{\tilde r}f_{\tilde r}
+4{\rm i}\,{\rm e}^{\mathscr{B-E}}f_{t}
-3f_{\tilde r}\Big\},
\label{K=1:piN:e2}
\end{align}
and then rewriting Eq.~\eqref{K=1:piN:e2} using Eq.~\eqref{K=1:pi:e}, we obtain
\begin{align}
\pi_{\tilde r}^{Z}&={\rm i}a\,\partial_{t}\pi_{\tilde r}
+{\rm e}^{\mathscr{A}}\Big\{-\tilde r\partial_{\tilde r}[{\rm e}^{-\mathscr{A}}(\pi_{\tilde r}-{\rm i}\,{\rm e}^{\mathscr{D}}f_{\tilde r})]+{\rm e}^{-\mathscr{A}}(\pi_{\tilde r}-{\rm i}\,{\rm e}^{\mathscr{D}}f_{\tilde r})
-4{\rm i}\,{\rm e}^{\mathscr{B-E}}\partial_{t}f_{t}
-\partial_{\tilde r}\left[4{\rm i}\,{\rm e}^{\mathscr{C-E}}f_{\tilde r}\right]\Big\}\notag\\
&\quad-{\rm i}\,{\rm e}^{\mathscr{D}}\Big\{
4{\rm i}\,{\rm e}^{-\mathscr{E}}\left(\partial_{\tilde r}\pi_{\tilde r}+{\rm e}^{\mathscr{B}}f_{t}\right)
-2f_{\tilde r}\Big\}.
\label{K=1:piN:e3}
\end{align}
Applying Eq.~\eqref{K=1:eom1}, we eliminate $f_{t}$ from Eq.~\eqref{K=1:piN:e3}:
\begin{align}
\pi_{\tilde r}^{Z}&={\rm i}a\,\partial_{t}\pi_{\tilde r}
\notag\\
&\quad+{\rm e}^{\mathscr{A}}\Big\{-\tilde r\partial_{\tilde r}\left[{\rm e}^{-\mathscr{A}}(\pi_{\tilde r}-{\rm i}{\rm e}^{\mathscr{D}}f_{\tilde r})\right]+{\rm e}^{-\mathscr{A}}(\pi_{\tilde r}-{\rm i}\,{\rm e}^{\mathscr{D}}f_{\tilde r})
-4{\rm i}\,{\rm e}^{-\mathscr{E}}\partial_{t}(-\partial_{\tilde r}\pi_{\tilde r}+{\rm i}\,{\rm e}^{\mathscr{E}}f_{\tilde r})
-\partial_{\tilde r}\left[4{\rm i}\,{\rm e}^{\mathscr{C-E}}f_{\tilde r}\right]\Big\}\notag\\
&\quad+6{\rm i}\,{\rm e}^{\mathscr{D}}f_{\tilde r}.
\label{K=1:piN:e4}
\end{align}
Applying Eq.~\eqref{K=1:eom2}, we eliminate $\partial_{t}f_{\tilde r}$ from Eq.~\eqref{K=1:piN:e4}:
\begin{align}
\pi_{\tilde r}^{Z}&={\rm i}a\,\partial_{t}\pi_{\tilde r}
\notag\\
&\quad+{\rm e}^{\mathscr{A}}\Big\{-\tilde r\partial_{\tilde r}({\rm e}^{-\mathscr{A}}\pi_{\tilde r})
+5{\rm e}^{-\mathscr{A}}\pi_{\tilde r}
+4{\rm i}\,{\rm e}^{-\mathscr{E}}\partial_{\tilde r}\left[\partial_{t}\pi_{\tilde r}
+\left({\rm e}^{2\mathscr{E-B}}-{\rm e}^{\mathscr{C}}+{\rm e}^{2\mathscr{D-A}}\right)f_{\tilde r}
+{\rm i}\,{\rm e}^{\mathscr{E-B}}\partial_{\tilde r}\pi_{\tilde r}
\right]
\Big\}.
\label{K=1:piN:e5}
\end{align}
Finally, applying Eq.~\eqref{K=1:eom3} to the last term of Eq.~\eqref{K=1:piN:e5}, we arrive at Eq.~\eqref{K=1:result}.

\section{Schwarzschild limit}
\label{app:C}

We take the Schwarzschild limit of the $(J,K,M)=(0,0,0)$, $(J,K,M)=(0,1,0)$ and $K=J+2$ modes we dealt with in Section~\ref{sec:III}, and we see the consistency with the vector and tensor harmonics $\mathbb{V}$ and $\mathbb{T}$ in Sec.~\ref{sec:II}, which can be understood well from the perspective of Killing vector fields.

\subsection{Killing vector fields and spherical harmonics on $S^{3}$}
\label{app:C1}

Let $K$ represent an arbitrary Killing vector field on $S^{3}$, obeying the Killing equation $\hat D_{i}K_{j}+\hat D_{j}K_{i}=0$ and hence divergenceless: $\hat D^{i}K_{i}=0$.
We have mentioned that the Riemann curvature tensor on $S^{3}$ is given in Eq.~\eqref{const}, and the Ricci tensor on $S^{3}$ is given as $\hat R_{ij}=2\gamma_{ij}$.
From the above properties of the Killing vector fields and the Ricci tensor, we have that $K$ is an eigenvector of the connection Laplacian $\hat\triangle=\hat D^{i}\hat D_{i}$,
\begin{align}
\hat\triangle K_{i}=\hat D^{j}\hat D_{j}K_{i}
&=-\hat D^{j}\hat D_{i}K_{j}\notag\\
&=-[\hat D^{j},\hat D_{i}]K_{j}-\hat D_{i}\hat D^{j}K_{j}
\notag\\
&=-\hat R^{k}{}_{i}K_{k}\notag\\
&=-2K_{i}.
\end{align}
Therefore, an arbitrary Killing vector field on $S^{3}$ is an eigenvector of the connection Laplacian on $S^{3}$ with eigenvalue $-2$, and it must be some linear combination of the three classes of vector harmonics, which have been defined in Eqs.~\eqref{V0}--\eqref{V2} based on Ref.~\cite{Lindblom:2017maa}.
Here, let us show the eigenvalue equations and the divergence of $\mathbb{V}^{(0)}$, $\mathbb{V}^{(1)}$ and $\mathbb{V}^{(2)}$ labeled by $(\underline{k},\underline{\ell},\underline{m})$ as follows,~\cite{Lindblom:2017maa}
\begin{subequations}
\begin{align}
\hat\triangle\mathbb{V}^{(0)}&=\Big(2-\underline{k}(\underline{k}+2)\Big)\mathbb{V}^{(0)},\\
\hat\triangle\mathbb{V}^{(1)}&=\Big(1-\underline{k}(\underline{k}+2)\Big)\mathbb{V}^{(1)},\\
\hat\triangle\mathbb{V}^{(2)}&=\Big(1-\underline{k}(\underline{k}+2)\Big)\mathbb{V}^{(2)},
\end{align}
\end{subequations}
and
\begin{subequations}
\begin{align}
\hat D\cdot\mathbb{V}^{(0)}&=-\sqrt{\underline{k}(\underline{k}+2)}\mathbb{V}^{(0)},\\
\hat D\cdot\mathbb{V}^{(1)}&=0,\\
\hat D\cdot\mathbb{V}^{(2)}&=0.
\end{align}
\end{subequations}
We can identify Killing vector fields as the linear combinations of vector harmonics from the above equations.
First, from the eigenvalue equations, the vector harmonics whose eigenvalue is $-2$ are only $\mathbb{V}^{(1)}$ and $\mathbb{V}^{(2)}$ with $\underline{k}=1$.
Secondly, from the equations on the divergence, both of $\mathbb{V}^{(1)}$ and $\mathbb{V}^{(2)}$ are divergenceless, which is consistent with the divergenceless property of Killing vector fields.
Thirdly, there are six independent vector harmonics $\mathbb{V}^{(1)}$ and $\mathbb{V}^{(2)}$ with $\underline{k}=1$, which is consistent with the number of the independent Killing vector fields on $S^{3}$; we count two vector harmonics for each of $(\underline{\ell},\underline{m})=(1,\pm1), (1,0)$, and 
$\mathbb{V}^{(1)}$ and $\mathbb{V}^{(2)}$ with $(\underline{\ell},\underline{m})=(0,0)$ vanish 
because the scalar harmonics with $(\underline{\ell},\underline{m})=(0,0)$ is given as $\mathbb{S}\propto\cos\chi$.
We recall that the classes of $\mathbb{V}^{(1)}$ and $\mathbb{V}^{(2)}$ are mapped to each other by the operator $\hat \star\hat d$ as
\begin{equation}
\hat \star\hat d\,\mathbb{V}^{(1)}{}=(\underline{k}+1)\mathbb{V}^{(2)},\quad
\hat \star\hat d\,\mathbb{V}^{(2)}{}=(\underline{k}+1)\mathbb{V}^{(1)}.
\label{V1V2}
\end{equation}
We have mentioned in Section.~\ref{sec:4A} that the covector fields $\bar\sigma^{i}$ on $S^{3}$ obey $\hat d\bar\sigma^{i}=-\epsilon^{i}{}_{jk}\bar\sigma^{j}\wedge \bar\sigma^{k}$, and we have also presented the expression~\eqref{gamma:sigma} of the metric $\gamma$ on $S^{3}$ in terms of $\bar\sigma^{i}$ in Section.~\ref{sec:4B}.
The operator $\hat \star\hat d$ acts on $\bar\sigma^{i}$ as
\begin{equation}
\hat \star\hat d\bar\sigma^{i}=-\epsilon^{i}{}_{jk}\epsilon^{jk}{}_{l}\bar\sigma^{l}=-2\bar\sigma^{i}.
\label{*dsigma}
\end{equation}
Comparing Eqs.~\eqref{V1V2} and \eqref{*dsigma}, we have that the three covector fields $\bar\sigma^{i}$ must be the linear combinations of the three vectors $\mathbb{V}^{(1)}-\mathbb{V}^{(2)}$ with $(\underline{k},\underline{\ell},\underline{m})=(1,1,\pm 1), (1,1,0)$.
Note that $(S^{3},\gamma)$ also admits the other three Killing vector fields $\xi^{i}$, and $\xi^{i}$ correspond to the three vectors $\mathbb{V}^{(1)}+\mathbb{V}^{(2)}$ with $(\underline{k},\underline{\ell},\underline{m})=(1,1,\pm 1), (1,1,0)$; the interested reader may refer to Ref.~\cite{Achour:2015zpa}.

Let us move on to the relation between the Killing vector fields and the tensor harmonics on $S^{3}$.
For convenience, we denote the symmetry operator as $\hat{\cal S}$, where we recall that the symmetry operator  $\hat{\cal S}$ maps a rank-2 symmetric tensor $T$ on $S^{3}$ to $\hat{\cal S}\,T_{ij}=2\hat\varepsilon_{(i}{}^{kl}\hat D_{|k|}T_{j)l}$.
We have that $\hat{\cal S}$ acts on the symmetric tensor product $\bar\sigma^{i}\bar\sigma^{j}=\frac{1}{2}(\bar\sigma^{i}\otimes\bar\sigma^{j}+\bar\sigma^{j}\otimes\bar\sigma^{i})$ of $\bar\sigma^{i}$ as
\begin{equation}
\hat{\cal S}\left[\bar\sigma^{i}\bar\sigma^{j}\right]
=-2\delta^{ij}(\bar\sigma^{1}\bar\sigma^{1}+\bar\sigma^{2}\bar\sigma^{2}+\bar\sigma^{3}\bar\sigma^{3})+6\bar\sigma^{i}\bar\sigma^{j}.
\label{Sss}
\end{equation}
From the above law of the map of $\bar\sigma^{i}\bar\sigma^{j}$ by the symmetry operator $\hat{\cal S}$, 
we obtain the six eigentensors of $\hat{\cal S}$ with eigenvalue $6$, including one linearly dependent tensor:
\begin{equation}
-2\bar\sigma^{i}\bar\sigma^{i}+\sum_{j\neq i}\bar\sigma^{j}\bar\sigma^{j}\quad(\text{not summed over $i$}),\qquad
\bar\sigma^{i}\bar\sigma^{j}\quad(i\neq j).
\label{ss}
\end{equation}
Let us uniformly denote these six tensors as $\Sigma_{ij}$ for convenience, and $\Sigma_{ij}$ are all found to be transverse and traceless, i.e. $\Sigma_{ij}$ satisfy $\Sigma^{i}{}_{i}=0$ and $\hat D^{i}\Sigma_{ij}=0$.
Note that the formula $\hat D\otimes\bar\sigma^{i}(=:\hat D_{j}(\bar\sigma^{i}{})_{k})=-\frac{1}{2}\epsilon^{i}{}_{jk}\bar\sigma^{j}\wedge \bar\sigma^{k}$ is useful to show the transverse property of $\Sigma_{ij}$.
Considered with these properties of $\Sigma_{ij}$, acting the symmetry operator $\hat{\cal S}$ on $\Sigma_{ij}$ twice leads to
\begin{equation}
\hat{\cal S}^{2}\Sigma_{ij}=4(3-\hat\triangle)\Sigma_{ij},
\end{equation}
and therefore $\Sigma_{ij}$ are found to be the eigentensors of the connection Laplacian $\hat\triangle$ with eigenvalue $-6$.
Then, we identify $\Sigma_{ij}$ in terms of the six classes of tensor harmonics on $S^{3}$, which have been defined in Eqs.~\eqref{T0}--\eqref{T5}.
The eigenvalue equations of $\hat\triangle$ and the divergences of the six classes of tensor harmonics are~\cite{Lindblom:2017maa}
\begin{subequations}
\begin{align}
\hat\triangle\mathbb{T}^{(0)}&=-\underline{k}(\underline{k}+2)\mathbb{T}^{(0)},\\
\hat\triangle\mathbb{T}^{(1)}&=\Big(5-\underline{k}(\underline{k}+2)\Big)\mathbb{T}^{(1)},\\
\hat\triangle\mathbb{T}^{(2)}&=\Big(5-\underline{k}(\underline{k}+2)\Big)\mathbb{T}^{(2)},\\
\hat\triangle\mathbb{T}^{(3)}&=\Big(6-\underline{k}(\underline{k}+2)\Big)\mathbb{T}^{(3)},\\
\hat\triangle\mathbb{T}^{(4)}&=\Big(2-\underline{k}(\underline{k}+2)\Big)\mathbb{T}^{(4)},\\
\hat\triangle\mathbb{T}^{(5)}&=\Big(2-\underline{k}(\underline{k}+2)\Big)\mathbb{T}^{(5)},
\end{align}
\end{subequations}
and
\begin{subequations}
\begin{align}
\hat D\cdot\mathbb{T}^{(0)}&=\frac{\sqrt{\underline{k}(\underline{k}+2)}}{\sqrt{3}}\,\mathbb{V}^{(0)},\\
\hat D\cdot\mathbb{T}^{(1)}&=-\frac{\sqrt{(\underline{k}-1)(\underline{k}+3)}}{\sqrt{2}}\,\mathbb{V}^{(1)},\\
\hat D\cdot\mathbb{T}^{(2)}&=-\frac{\sqrt{(\underline{k}-1)(\underline{k}+3)}}{\sqrt{2}}\,\mathbb{V}^{(2)},\\
\hat D\cdot\mathbb{T}^{(3)}&=-\frac{\sqrt{2(\underline{k}-1)(\underline{k}+3)}}{\sqrt{3}}\,\mathbb{V}^{(0)},\\
\hat D\cdot\mathbb{T}^{(4)}&=0,\\
\hat D\cdot\mathbb{T}^{(5)}&=0.
\end{align}
\end{subequations}
On the other hand, we have shown that $\Sigma_{ij}$ satisfy $\hat\triangle\Sigma=-6\Sigma$ and $\hat D\cdot\Sigma=0$, and therefore $\Sigma_{ij}$ must be the linear combination of the tensor harmonics $\mathbb{T}^{(4)}$ and $\mathbb{T}^{(5)}$ with $\underline{k}=2$.
From Eq.~\eqref{T5}, $\mathbb{T}^{(4)}$ and $\mathbb{T}^{(5)}$ with the same label $(\underline{k},\underline{\ell},\underline{m})$ are mapped to each other as
\begin{equation}
\hat{\cal S}\,\mathbb{T}^{(4)}=2(\underline{k}+1)\mathbb{T}^{(5)},\quad \hat{\cal S}\,\mathbb{T}^{(5)}=2(\underline{k}+1)\mathbb{T}^{(4)},
\end{equation}
and, according to Ref.~\cite{Lindblom:2017maa}, only $\underline{\ell}=2$ gives non-zero $\mathbb{T}^{(4)}$ and $\mathbb{T}^{(5)}$ with $\underline{k}=2$.
Therefore, $\Sigma_{ij}$ must correspond to the combination $\mathbb{T}^{(4)}+\mathbb{T}^{(5)}$, labeled with $(\underline{k},\underline{\ell},\underline{m})=(2,2,\pm 2),(2,2,\pm 1),(2,2,\pm 0)$, where we remark that the number of the allowed labels $(\underline{k},\underline{\ell},\underline{m})$ is consistent with that of linearly independent rank-2 symmetric traceless tensors $\Sigma_{ij}$, presented in Eq.~\eqref{ss}.
Note that the combination $\mathbb{T}^{(4)}-\mathbb{T}^{(5)}$ correspond to the tensors constructed with the Killing vector fields $\xi^{i}$ as well as the aforementioned investigation of the Killing vector fields in terms of the vector harmonics.

\subsection{$(J,K,M)=(0,0,0)$ mode}

In the Schwarzschild limit $a=0$, the image $Z_{\lambda\rho}$ of the rank-2 symmetric tensor $h_{\lambda\rho}$ with $(J,K,M)=(0,0,0)$ by the symmetry operator, given in Eqs.~\eqref{K=0:gaugeN:tt}--\eqref{K=0:gaugeN:+-}, reduces to
\begin{equation}
Z_{\tilde r 3}=2h_{\tilde r 3},\quad Z_{33}=-4h_{+-},\quad Z_{+-}=2h_{+-},
\end{equation}
and Eq.~\eqref{K=0:Gr3}, one of the perturbation equations for $h_{\lambda\rho}$, leads to $Z_{\tilde r 3}=h_{\tilde r 3}=0$.
Therefore, we arrive at
\begin{align}
Z
&=2h_{+-}\left(2\bar\sigma^{+}\bar\sigma^{-}-2\bar\sigma^{3}\bar\sigma^{3}\right)
\notag\\
&=2 h_{+-}\left(\bar\sigma^{1}\bar\sigma^{1}+\bar\sigma^{2}\bar\sigma^{2}-2\bar\sigma^{3}\bar\sigma^{3}\right),
\end{align}
that is to say, the $(J,K,M)=(0,0,0)$ mode is included in the family of $\mathbb{T}^{(4)}+\mathbb{T}^{(5)}$ with $(\underline{k},\underline{\ell})=(2,2)$.

\subsection{$(J,K,M)=(0,1,0)$ mode}
\label{app:C3}

In the Schwarzschild limit $a=0$, the image $Z_{\lambda\rho}$ of the rank-2 symmetric tensor $h_{\lambda\rho}$ with $(J,K,M)=(0,1,0)$ by the symmetry operator, given in Eqs.~\eqref{K=1:Zt+}--\eqref{K=1:Z+3}, reduces to
\begin{equation}
Z_{t+}=2h_{t+},\quad
Z_{\tilde r +}=2h_{\tilde r +},\quad
Z_{+3}=6h_{+3},
\end{equation}
and the perturbation equation yields that $h_{t+}=h_{\tilde r +}=0$, which follows from the observation on the so-called exceptional modes of vector harmonics (see Ref.~\cite{Ishibashi:2011ws}).
To summarize, we arrive at
\begin{align}
Z
&=6\left(2h_{+3}\bar\sigma^{+}\bar\sigma^{3}+2h_{-3}\bar\sigma^{-}\bar\sigma^{3}\right)
\notag\\
&=6\left(2h_{13}\bar\sigma^{1}\bar\sigma^{3}+2h_{23}\bar\sigma^{2}\bar\sigma^{3}\right),
\end{align}
where
\begin{equation}
h_{13}=\frac{1}{\sqrt{2}}\left(h_{+3}+h_{-3}\right),\quad
h_{23}=\frac{{\rm i}}{\sqrt{2}}\left(h_{+3}-h_{-3}\right).
\end{equation}
Therefore, we have that the $(J,K,M)=(0,1,0)$ mode is included in the family of $\mathbb{T}^{(4)}+\mathbb{T}^{(5)}$ with $(\underline{k},\underline{\ell})=(2,2)$.

\subsection{$K=J+2$ modes}

From Eq.~\eqref{K=J+2:Z}, in the Schwarzschild limit $a=0$, the rank-2 symmetric tensor $h_{\lambda\rho}$ with $K=J+2$, given by the form~\eqref{K=J+2:h}, is mapped by the symmetry operator to
\begin{equation}
Z=\hat{\cal S}h=2(2J+3)h.
\end{equation}
Let us also evaluate the eigenvalue of $h$ with respect to the connection Laplacian $\hat\triangle$ by applying the formula $\hat D\otimes\bar\sigma^{\pm}(=\hat D_{i}(\bar\sigma^{\pm})_{j})=\pm {\rm i}\bar\sigma^{\pm}\wedge\bar\sigma^{3}$ and the relation $\hat\triangle=-4W^{2}$ between the Laplace-Beltrami operator $\hat\triangle$ and the Casimir operator $W^{2}$.
Note that we denote the Wigner $\mathbb{D}$ functions as $\mathbb{D}_{\pm J}:=\mathbb{D}^{J}{}_{K=\pm J\,M=0}$ for convenience.
We obtain the following eigenvalue equation for $h_{\lambda\rho}$:
\begin{align}
\hat\triangle h&=\hat\triangle\Big[\sum_{\pm}h_{\pm\pm}(t,\tilde r)\mathbb{D}_{\pm J}\bar\sigma^{\pm}\bar\sigma^{\pm}\Big]\notag\\
&=\sum_{\pm}h_{\pm\pm}(t,\tilde r)\left\{
(\hat\triangle \mathbb{D}_{\pm J})\bar\sigma^{\pm}\bar\sigma^{\pm}+2\mathbb{D}_{\pm J}(\hat\triangle\bar\sigma^{\pm})\bar\sigma^{\pm}
+2\left[2(\hat D_{i}\,\mathbb{D}_{\pm J})(\hat D^{i}\bar\sigma^{\pm})\bar\sigma^{\pm}+\mathbb{D}_{\pm J}(\hat D^{i}\bar\sigma^{\pm})(\hat D^{i}\bar\sigma^{\pm})\right]\right\}\notag\\
&=\sum_{\pm}h_{\pm\pm}(t,\tilde r)\left\{
(-4W^{2}\mathbb{D}_{\pm J})\bar\sigma^{\pm}\bar\sigma^{\pm}-4\mathbb{D}_{\pm J}\bar\sigma^{\pm}\bar\sigma^{\pm}
+2\left[\pm 2{\rm i}(-\sqrt{2}\,{\rm i}\,W_{\pm}\mathbb{D}_{\pm J})\bar\sigma^{3}\bar\sigma^{\pm}-\mathbb{D}_{\pm J}\bar\sigma^{\pm}\bar\sigma^{\pm}\right]\right\}\notag\\
&=\sum_{\pm}h_{\pm\pm}(t,\tilde r)\left\{
-4J(J+2)\mathbb{D}_{\pm J}\bar\sigma^{\pm}\bar\sigma^{\pm}-4\mathbb{D}_{\pm J}\bar\sigma^{\pm}\bar\sigma^{\pm}
-2\mathbb{D}_{\pm J}\bar\sigma^{\pm}\bar\sigma^{\pm}\right\}\notag\\
&=\Big(2-(2J+2)(2J+4)\Big)h.
\end{align}
Here, by comparing the eigenvalue equations $\hat\triangle\mathbb{S}^{\underline{k'\ell' m'}}=-\underline{k'}(\underline{k'}+2)\mathbb{S}^{\underline{k'\ell' m'}}$ for the scalar harmonics $\mathbb{S}^{\underline{k'\ell' m'}}$ and $W^{2}\mathbb{D}^{J}_{KM}=J(J+1)\mathbb{D}^{J}_{KM}$ for the Wigner $\mathbb{D}$ function $\mathbb{D}^{J}_{KM}$, we have that the labels $\underline{k'}$ and $J$ are related by $\underline{k'}=2J$, and, in addition, we define another label $\underline{k}:=\underline{k'}+2$ in terms of the tensor harmonics $\mathbb{T}$.
The above equations are rewritten as
\begin{equation}
\hat{\cal S}h=2\left(\underline{k}+1\right)h,\quad
\hat\triangle h=\left(2-\underline{k}(\underline{k}+2)\right)h.
\end{equation}
Therefore, the $K=J+2$ mode is included in the family of $\mathbb{T}^{(4)}+\mathbb{T}^{(5)}$ labeled by $\underline{k}=\underline{k'}+2$.

\end{document}